\documentclass[reprint,notitlepage,superscriptaddress,preprintnumbers,nofootinbib,nobibnotes,amsmath,amssymb,aps,twocolumn]{revtex4-1}
\usepackage{graphicx,grffile}
\usepackage{dcolumn}
\usepackage{bm}
\usepackage{hyperref}
\usepackage{aas_macros,braket}

\makeatletter
\def\lst@lettertrue{\let\lst@ifletter\iffalse}
\makeatother

\begin{document}
\title{Detecting higher spin fields through statistical anisotropy in the CMB bispectrum}
\author{Gabriele Franciolini}
\affiliation{D\'epartement de Physique Th\'eorique and Centre for Astroparticle Physics (CAP), Universit\'e de Gen\`eve, 24 quai E. Ansermet, CH-1211 Geneva, Switzerland}
\author{Alex Kehagias} 
\affiliation{Physics Division, National Technical University of Athens, 15780 Zografou Campus, Athens, Greece}
\author{Antonio Riotto}
\affiliation{D\'epartement de Physique Th\'eorique and Centre for Astroparticle Physics (CAP), Universit\'e de Gen\`eve, 24 quai E. Ansermet, CH-1211 Geneva, Switzerland}
\author{Maresuke Shiraishi}
\affiliation{Department of General Education, National Institute of Technology, Kagawa College, 355 Chokushi-cho, Takamatsu, Kagawa 761-8058, Japan}

\date{\today}


\begin{abstract}

  Inflation may provide a suitable collider to probe physics at very high energies. In this paper we investigate the impact on the CMB bispectrum of higher spin fields which are long-lived on super-Hubble scales, e.g. partially massless higher spin fields. We show that distinctive statistical anisotropic signals on the CMB three-point correlator are induced and we investigate their detectability. 
 
\end{abstract}

\maketitle


\section{Introduction}

Inflation \cite{Lyth:1998xn} offers the best paradigm explaining the dynamics of the early universe and a plethora of cosmological observations performed to date. In particular, it provides a mechanism  to generate the cosmological perturbations responsible for the cosmic microwave background (CMB) anisotropies we observe today. The particle content of minimal inflationary models is composed by the gravitational degrees of freedom (\textsl{d.o.f.}) and a slow-rolling scalar field $\phi$ causing the accelerated expansion. The cosmological perturbations are then interpreted as quantum fluctuations of the inflaton field which are stretched to super-Hubble scales. In multi-field models of inflation it is assumed that an higher number of fields are responsible both for the dynamics of the inflationary Friedmann-Lema\^itre-Robertson-Walker background as well as the features of the cosmological perturbations. Indeed, constraining even further the inflationary epoch and determining the particle content of the early universe which could have left specific signatures in the primordial curvature perturbations is undoubtedly one of the main challenges of modern cosmology.
 
 One way to pursue this task is presented by the possible detection of primordial non-Gaussianity (NG). At this point, NG has been extensively studied (for a review see Refs.~\cite{Bartolo:2004if,Chen:2010xka}) and constrained by the latest observations \cite{Ade:2015ava,Ade:2015lrj}. At the same time, the presence of anisotropies and/or inhomogeneities could modify NG by providing, in turn, possible signatures of extra \textsl{d.o.f.} present during the inflationary stage. In other words, there could be, for example, peculiar signatures on NG that would not be produced by inflaton self-interactions but which were due to new \textsl{d.o.f.} dynamically excited during the inflationary era.

\begin{figure*}[t]
  \begin{tabular}{c}
    \begin{minipage}{1.\hsize}
  \begin{center}
    \includegraphics[width = 1.\textwidth]{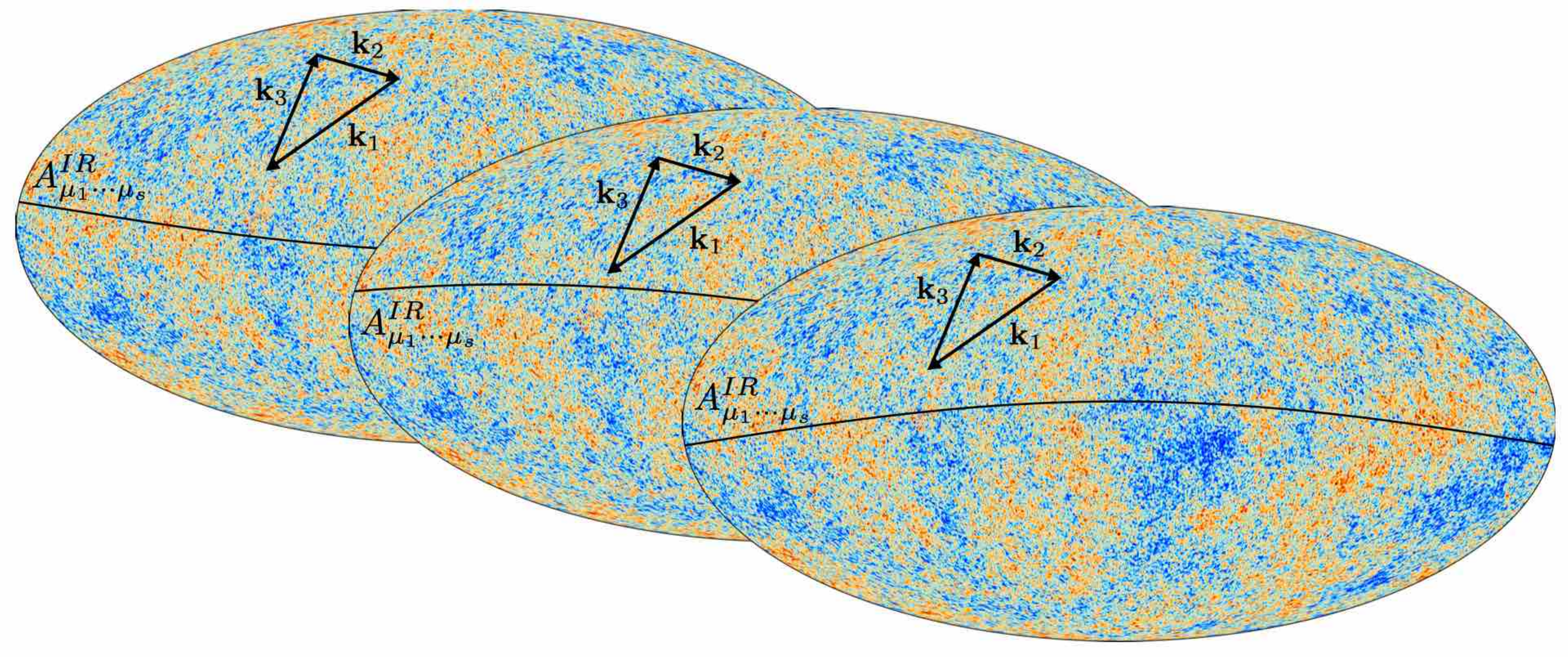}
  \end{center}
\end{minipage}
\end{tabular}
  \caption{A schematic view of various realizations of the first $N-N_k$ e-folds of inflation where each of the unsuppressed IR super-Hubble higher spin modes $A_{\mu_1 \cdots \mu_s }^{IR}$ behaves as non-trivial background. The cosmological correlators are sensitive to the value the IR modes assume in a  \textsl{single} realization of the ensemble of possible universes.}\label{fig:CMB} 
\end{figure*}
 
From a particle physics perspective, since the inflationary epoch is characterized by typical energies much larger than the electroweak scale which experiments at colliders have been able to reach so far, it provides the best playground to test physics at high energies \cite{Chen:2009we,Chen:2009zp,Arkani-Hamed:2015bza}. In general, it is interesting to investigate how any supposed new particle content could have left imprints on the cosmological observables. In particular, one can find a review of the effects of extra light scalar fields on NG in Ref.~\cite{Byrnes:2010em} while other cases were studied where scalar fields with masses larger than the Hubble scale ($m\geq H$) were still being dynamically excited during inflation (e.g. quasi-single fields models \cite{Chen:2009we,Chen:2009zp,Baumann:2011nk,Noumi2013,Kehagias:2015jha,Dimastrogiovanni:2015pla}).
Furthermore, it was shown \cite{Barnaby:2012tk,Bartolo:2012sd,Shiraishi:2013vja} that a vector field, whose interactions are described by a coupling term of the form $I(\phi) F^2$ in the action, gives rise to a non-trivial angular dependence of the curvature bispectrum, see also Ref.~\cite{Bartolo:2015dga}. Analogous angular structures can be found in solid inflation models \cite{Endlich:2012pz} as well as models with primordial magnetic fields \cite{Shiraishi:2012rm,Shiraishi:2012sn}. Finally, it was suggested that correlators of comoving curvature perturbations could inherit NG features due to interactions of massive higher spin particles \cite{Arkani-Hamed:2015bza,Lee:2016vti,Baumann:2017jvh,Franciolini:2017ktv}.

There is a particularly interesting perspective of under which one can characterize the possible effects of higher spin field on cosmological correlators. This is provided by the so called dS/CFT$_3$ correspondence \cite{Strominger:2001pn}. It is widely accepted that inflation can be described as a phase of (quasi-)de Sitter spacetime. During this period, the isometries of spacetime form an (approximate) SO(4,1) group which can be identified with the conformal group of a CFT$_3$ acting on perturbations extending on scales larger than the Hubble radius. The higher spin field evolution can be written as $A_{i_1 \cdots i_s}(\vec{x} , \tau) = (-\tau )^{\Delta-s}A_{i_1 \cdots i_s}(\vec{x})$ with $\Delta= {3}/{2} - \sqrt{(s-{1}/{2})^2-{m^2}/{H^2}}$ when the conformal time $\tau$ goes to zero, i.e. on the boundary of dS. Using the isometries of the de Sitter spacetime it is possible to show that the masses and the spins of the higher spin fields must be related through what is called the Higuchi bound \cite{Higuchi:1986py}
\begin{equation}
m^2 > s(s-1) H^2, \label{ea:Hig}
\end{equation}
in order to comply to the unitarity bounds. This requires $\Delta >1$ and therefore the typical amplitudes of the higher spin fluctuations must decay faster than $e^{-Ht}$ on super-Hubble scales \cite{Arkani-Hamed:2015bza,Bordin:2016ruc,Kehagias:2017rpe}, suppressing their contribution to the bispectrum. 

The technical difficulty in generating long-lived spin-$s$ perturbations can be  overcome in two possible ways: one  is provided by the introduction of an apposite coupling between the higher spins and the inflaton \cite{Kehagias:2017cym}, in analogy with the aforementioned vector case; alternatively, one can restrict the analysis to partially massless higher spin fields \cite{Deser:1983mm,Deser:1983tm,Deser:2001pe,Deser:2001us,Deser:2003gw} which, by having particular values of their masses, present unsuppressed perturbations on super-Hubble scales for some helicities states \cite{Baumann:2017jvh,Franciolini:2017ktv}, i.e. states with $\Delta =0$.

A peculiar phenomenon happens in theories where higher spin long-lived perturbations are present, namely the spontaneous generation of a non-trivial expectation value for those fields leading to a statistical anisotropy in the correlators, see Fig.~\ref{fig:CMB}. This phenomenon  was first described  in theories where the U(1) field couples to the inflaton field, but the mechanism can be generalized to both higher spins coupled to inflaton and partially massless higher spins. Indeed, as was shown in Refs.~\cite{Ratra:1991bn,Martin:2007ue,Demozzi:2009fu,Bartolo:2012sd,Biagetti:2013qqa}, the spin-1 perturbations can be rendered constant on super-Hubble scales by choosing a suitable coupling to the inflaton $I(\phi)$ and constant energy density is stored into the infrared (IR) quantum mechanical fluctuations developed in the previous stages of inflation which has exited the Hubble radius. Those ``frozen" perturbations present a typical amplitude proportional to the square root of the variance of the field. In the later stages of inflation those constant modes act, in turn, as a classical background creating a non-vanishing vacuum expectation value of the vector field corresponding to a \textsl{single} realization of the first $(N-N_k)$ e-folds of inflation,%
\footnote{$N$ is the total number of e-folds while $N_k$ is the number of e-folds before the end of inflation corresponding to the moment when the comoving scale $k^{-1}$ exit the comoving Hubble horizon.}
leading necessarily to a  preferential direction and therefore breaking of  the isotropy. As a result, the power spectrum and the higher-order correlators of the curvature perturbations resulting from interactions with the U(1) field present a distinct pattern responsible for the generation of a statistical anisotropy, see also Refs.~\cite{Watanabe:2010fh,Soda:2012zm}. Generalizations of this mechanism for higher spin particles were studied in Refs. \cite{Kehagias:2017cym,Bartolo:2017sbu,Franciolini:2017ktv} within the specific model where massive higher spin fields are made effectively massless during inflation by suitable couplings to the inflaton field. In particular, it was shown there that the statistical anisotropy imprints in the CMB and galaxy power spectra from higher spin fields can be detected at the level of a few percents. 

In this paper we take a further step towards the quantitative analysis of the statistical anisotropies induced by higher spin fields and analyze their imprint on the primordial curvature bispectrum. More in general, we study the template
\begin{eqnarray}
  \begin{aligned}
    \Braket{\prod_{i=1}^3 \zeta_{\vec{k}_i}}  &= (2\pi)^3 \delta^{(3)}\left(\sum_{i=1}^3 \vec{k}_i \right) \sum_{n} c_n {\cal B}_{k_1 k_2 k_3}^n, \\
    {\cal B}_{k_1 k_2 k_3}^n &\equiv P_n(\hat{k}_1 \cdot \hat{k}_2) P_\zeta(k_1)P_\zeta(k_2) + 2~{\rm perm}, \label{eq:zeta3_cn}
  \end{aligned}
\end{eqnarray}
where $P_n$ is the Legendre polynomials.  The theoretical predictions for the form of the coefficients $c_n$ in the case of spinning particles coupled to the inflaton and for partially massless fields studied in detail in Ref.~\cite{Franciolini:2017ktv} are reviewed in Sec.~\ref{sec:prim}. As we will see, higher spin particles induce $s+1$ nonvanishing coefficients with $n$ even: $c_0, c_2, \cdots, c_{2s-2}$ and $c_{2s}$.

Recently, the signatures of $c_n$ in diverse cosmic observables have been widely investigated (see e.g., Refs.~\cite{Shiraishi:2013vja,Shiraishi:2016hjd} for CMB, Refs.~\cite{Byun:2014cea,Raccanelli:2015oma,Assassi:2015jqa,Schmidt:2015xka,Assassi:2015fma,Chisari:2016xki,MoradinezhadDizgah:2017szk,MoradinezhadDizgah:2018ssw} for the large-scale structure (LSS), and Ref.~\cite{Munoz:2015eqa} for 21-cm fluctuations). Moreover, the observational bounds on $c_{0, 1, 2}$ have already been obtained via the CMB bispectrum estimation by the {\it Planck} team \cite{Ade:2013ydc,Ade:2015ava}. On the other hand, these works focus only on smaller $n$. In this paper, for the first time  the contributions due to nonvanishing $c_{n \geq 4}$ to the CMB bispectra are presented via the extension of the previous study \cite{Shiraishi:2013vja} discussing up to $n = 2$. The detection of the coefficients $c_n$ in the curvature bispectrum~\eqref{eq:zeta3_cn} could provide insight into the inflationary dynamics and particles content during inflation, e.g. determining whether higher spin fields were present in the early stages of the universe.

The paper is organized as follows. In Sec.~\ref{sec:prim} we review and the summarize the results contained in Ref.~\cite{Franciolini:2017ktv} giving in detail the theoretical prediction of the bispectrum in the presence of spinning particles while Sec.~\ref{sec:CMB_bis} is devoted to the detailed analysis and forecast for CMB experiments providing the sensibility to higher spin fields effects. In Sec.~\ref{sec:con} we present our conclusions.

\section{Anisotropic non-Gaussianities generated from higher spin fields}
\label{sec:prim}

In this section and for the convenience of the reader we summarize the results found in Ref.~\cite{Franciolini:2017ktv} computing the angular dependence of the statistical anisotropic curvature bispectrum due to higher spin fields  in the early universe, thus providing a prediction for the coefficients $c_n$ in the bispectrum parametrization~\eqref{eq:zeta3_cn}. As stated in the introduction, in order to have unsuppressed effects on the bispectrum from higher spins, one has to rely on either of the two mechanisms, namely by coupling the higher spins to the inflaton in a precise manner or by restricting the analysis to partially massless higher spin fields. The predictions in the two cases are different, as we shall see. 

We begin by reviewing the results for spin-1 fields coupled to the inflaton through an interaction term of the form:
\begin{equation}
\mathcal{L} = -\frac{I(\phi)}{4} F_{\mu\nu} F^{\mu\nu}.
\end{equation}
Working with gauge invariant quantities, one defines the ``electric field''%
\footnote{Even though it is customary to borrow the notations from the theory of electromagnetism, the spin-1 field is not necessarily identified with the standard model photon field.}
as $E_i = - a^{-2} \Braket{I^{1/2}} A_i' $ where the prime symbol denotes derivatives with respect to conformal time. Using the equations of motion, it is straightforward to check that for a constant $I$, the contribution of the electric field is suppressed as $E_i \propto a^{-2}$. On the other hand, if one assume to have a coupling with $\Braket{I^{1/2}}\propto a^{-2}$, the electric field is constant. From a dS/CFT$_3$ point of view this corresponds to super-Hubble electric perturbations with a vanishing conformal dimension $\Delta_E =0$. As a result, defining $\vec{E}_{c} = \vec{E}_{(0)}+ \vec{E}^{IR}   $ where one accounts for a possible background electric field $\vec{E}_{(0)}$ other than the one necessarily generated by the IR modes $\vec{E}^{IR}$, the inflaton-vector interaction lagrangian expanded in perturbations contains terms of the form $ \delta E_i E_c^i \zeta $ and $(\delta E_i )^2 \zeta$ which give rise to the following bispectrum
\begin{eqnarray}
  \Braket{\prod_{i=1}^3 \zeta_{\vec{k}_i}}'
  &=& g_1 | \vec{E}_{c} |^2 P_\zeta(k_1)P_\zeta(k_2) \nonumber \\
  && \left(1 - \cos^2\theta_{\hat{k}_1, {\hat{E_{c}}}} - \cos^2\theta_{\hat{k}_2, {\hat{E_{c}}}} \right. \nonumber \\
  && 
  \left. +\cos\theta_{\hat{k}_1, \hat{k}_2}\cos\theta_{\hat{k}_1, {\hat{E_{c}}}}\cos\theta_{\hat{k}_2, {\hat{E_{c}}}}\right) \nonumber \\
  && 
  + 2~{\rm perm} , \label{eq:spin1}
\end{eqnarray}
where $\cos\theta_{\hat{q}_1, \hat{q}_2} \equiv \hat{q}_1 \cdot \hat{q}_2$, the exact value of the coefficient $g_1$ can be found in Ref.~\cite{Bartolo:2012sd}, and we have used the standard notation of indicating the prime the correlators without the $(2\pi)^3$ and the momentum conservation Dirac delta.

Generalizing this result for higher spin fields, the authors of Ref.~\cite{Kehagias:2017cym} showed that there exists a suitable coupling rendering the higher spin perturbations constant on super-Hubble scales. Following a bottom-up approach starting from the equations of motion for the higher spin field, it was found that such a configuration can be achieved by a coupling of the form:
\begin{equation}
\mathcal{S} \supset \beta_s H^2 \int \frac{d \tau d^3x}{ H^4 \tau ^4} \exp{\left(I(\phi)\right)} A_{i_1\cdots i_s} A^{i_1\cdots i_s}.
\end{equation}
Indeed, the equation of motion for the dominant mode of helicity $s$ of the higher spin field is:%
\footnote{The mode with helicity $\lambda=s$ is dominant on super-Hubble scales since the conformal weight of helicity $\lambda$ modes is, in general, $\Delta_s = 1- \lambda - 1/ \alpha$.}
\begin{eqnarray}
  &&A_s'' - \left[ \frac{2(1-s)}{\tau } - I' \right]A_s' \nonumber \\
  && + \left[ k^2 + \frac{M_s^2 / H^2 + s^2 -4 s }{\tau^2 } + \frac{s(1 + \alpha)}{\tau} I' \right] A_s =0,
\end{eqnarray}
where we have defined $M_s^2 =(s+2) H^2 (\alpha - s\alpha -1 )/ \alpha  $, $I(\phi(\tau)) = \ln (-H \tau)/\alpha$ and $\alpha = 1/[2(1-s)]$. This configuration generates constant perturbations on super-Hubble scales for the canonically normalized field $\bar{A}_s =\exp{\left(I(\phi) / 2 \right)} A_s$.  Using the isometries of de Sitter spacetime%
\footnote{
The symmetry considerations used are exactly valid in multi-field models of inflation where the curvature perturbations are generated by spectator fields (e.g. the curvaton mechanism \cite{Lyth:2001nq}) and SO(1,4) is an exact symmetry of the action. In single-field models of inflation, special conformal symmetry is broken and we expect correction at leading order in the slow-roll parameters.
} and the Operator Product Expansion (OPE) technique, the form of the statistically anisotropic curvature bispectrum is found to be \cite{Franciolini:2017ktv}:
\begin{eqnarray} 
  \Braket{\prod_{i=1}^3 \zeta_{\vec{k}_i} }'
  &=& g_s \Braket{ \bar{A}^{i_1\cdots i_s}} \Braket{ \bar{A}^{j_1\cdots j_s}} P_\zeta(k_1)P_\zeta(k_2) \nonumber \\ 
&&  \Pi_{i_1\cdots i_s}^{\ell_1 \cdots \ell_s} (\vec{k}_1)\Pi_{j_1\cdots j_s}^{\ell_1 \cdots \ell_s} (\vec{k}_2)  +  2~{\rm perm},  \label{eq:bis-hs}
\end{eqnarray} 
where the projector tensor $\Pi_{i_1\cdots i_s}^{m_1 \cdots m_s} (\vec{k})$ is defined in terms of the sum of helicities of the higher spin polarization tensors, see Appendix~\ref{app:1}. We assume that the isotropy is broken by a constant unit vector $\hat{p}_{i}$ ($\hat{p} \cdot \hat{p} = 1$) and the background generated by the IR modes takes the form 
\begin{eqnarray}
&& \Braket{\bar{A}_{i_1 \cdots i_s}} = \bar{A_0} \left[\hat{p}_{i_1}\cdots \hat{p}_{i_s} \right. \nonumber \\
  && \qquad \left. -\frac{1}{2s-1} \left(\delta _{i_1 i_2} \hat{p}_{i_3} \cdots \hat{p}_{i_s} +  {\rm perm} \right)+ \cdots \right], \label{eq:vac}
\end{eqnarray}
where the ellipsis stands for terms with an higher number of Kronecker delta which are built by requiring Eq.~\eqref{eq:vac} to be totally symmetry and traceless. Those terms are omitted for simplicity since  they are not relevant in the following calculations. It is important to stress that it was assumed that the IR modes select only a single preferential direction identified by the constant unit vector $\hat{p}_{i}$. As already underlined in the introduction, we expect the typical amplitude of the background field to be proportional to the square root of the variance $A_0 \sim H \sqrt{N-N_k}$. It is straightforward to see that combining Eq.~\eqref{eq:bis-hs} with the formula for the vacuum expectation value \eqref{eq:vac} for spin-1 gives Eq. \eqref{eq:spin1} up to a normalization constant which the symmetry considerations combined with the OPE are unable to predict.

While the formula \eqref{eq:bis-hs} is the exact one, for the CMB analysis the isotropized form 
\begin{equation}
\int \frac{ d^2 \hat{p}}{4\pi} \Braket{\prod_{i=1}^3 \zeta_{\vec{k}_i} }
\end{equation}
 is introduced. 
Using the definitions of the projector tensors \eqref{eq:proj}, we expand the bispectrum formula and we isolate the term containing the vectorial dependences, which we call $I_s(\hat{p}, \vec{k}_1 , \vec{k}_2)$:
\begin{eqnarray}
I_s(\hat{p},\vec{k}_1,\vec{k}_2) &=& \Braket{ \bar{A}^{i_1\cdots i_s}}
\Braket{ \bar{A}^{j_1\cdots j_s}} \nonumber \\
&& \sum_{\lambda_1}   \epsilon_{\lambda_1, i_1\cdots i_s}(\vec{k}_1)
\epsilon_{\lambda_1}^{*\ell_1\cdots \ell_s}(\vec{k}_1) \nonumber \\
&&
\sum_{\lambda_2} \epsilon^*_{\lambda_2 ,j_1\cdots j_s}(\vec{k}_2) \epsilon_{\lambda_2}^{\ell_1\cdots \ell_s}(\vec{k}_2),
\end{eqnarray}
using the fact that the polarization tensor is traceless we obtain 
\begin{eqnarray}
I_s(\hat{p},\vec{k}_1,\vec{k}_2) &=& \bar{A}_0 ^2\, \hat{p}^{i_1} \cdots \hat{p}^{i_s}\hat{p}^{j_1}\cdots \hat{p}^{j_s}  \nonumber \\
&& \sum_{\lambda_1}   \epsilon_{\lambda_1, i_1\cdots i_s}(\vec{k}_1)
\epsilon_{\lambda_1}^{*\ell_1\cdots \ell_s}(\vec{k}_1) \nonumber \\
&&
\sum_{\lambda_2} \epsilon^*_{\lambda_2 ,j_1\cdots j_s}(\vec{k}_2) \epsilon_{\lambda_2}^{\ell_1\cdots \ell_s}(\vec{k}_2). \label{is}
\end{eqnarray}
Now we perform the angle average over all the possible directions $\hat{p}$ by computing%
\footnote{In performing the angle average we use the normalization
  \begin{equation}
    \int \frac{ d^2 \hat{p}}{4\pi}\, \hat{p}^i\hat{p}^j = \frac{1}{3}\delta^{ij},
  \end{equation}
and the identities in e.g. Ref.~\cite{Ee:2017jxx}.}
\begin{eqnarray}
  I_{s,{\rm av}}(\vec{k}_1,\vec{k}_2) &=& \int \frac{ d^2 \hat{p}}{4\pi} \, I_s(\hat{p},\vec{k}_1,\vec{k}_2) \nonumber \\
  &=&  I^0_s 
  \sum_{\lambda_1 \lambda_2} \left| \epsilon_{\lambda_1, i_1 \cdots i_s}(\vec{k}_1) \epsilon^*_{\lambda_2, i_1 \cdots i_s}(\vec{k}_2) \right|^2 \nonumber \\
  &=& I^0_s \, \Pi_{i_1\cdots i_s}^{\ell_1\cdots \ell_s}(\vec{k}_1) 
\Pi_{i_1\cdots i_s}^{\ell_1\cdots \ell_s}(\vec{k}_2) ,
\end{eqnarray}
where $I^0_s=\bar{A}_0 ^2 s!/(2s+1)!!$. The computation of the helicities sums is easily performed for $s=1$ where one finds:
  \begin{equation}
\Pi_{i}^{\ell}(\vec{k}_1) \Pi_{i}^{\ell}(\vec{k}_2) 
    = 1 + \cos^2\theta_{\hat{k}_1, \hat{k}_2} .
  \end{equation}
The generalization to spin-$s$ is done by adopting Eq.~\eqref{eq:app.1} to get:
\begin{eqnarray}
&& \Pi_{i_1\cdots i_s}^{\ell_1\cdots \ell_s}(\vec{k}_1) 
\Pi_{i_1\cdots i_s}^{\ell_1\cdots \ell_s}(\vec{k}_2) \nonumber \\ 
  &&\quad = \frac{1}{2}
    \left\{(1+\cos\theta_{\hat{k}_1, \hat{k}_2})^{2s} 
    +(1-\cos\theta_{\hat{k}_1, \hat{k}_2})^{2s} \right\}  ,
\end{eqnarray}
therefore we can write the angle averaged bispectrum as
\begin{eqnarray} 
  \Braket{\prod_{i=1}^3 \zeta_{\vec{k}_i} }'_{\text{av}}
  &=& \frac{1}{2} g_s I^0_s P_\zeta(k_1)P_\zeta(k_2) \nonumber \\ 
  &&  
  \left\{(1+\cos\theta_{\hat{k}_1, \hat{k}_2})^{2s}
 +(1-\cos\theta_{\hat{k}_1, \hat{k}_2})^{2s}\right\} \nonumber \\ 
   && + 2~{\rm perm}.
\end{eqnarray} 
Now we expand the result using the Legendre polynomials $P_n(\cos \theta)$ using the standard formulas:
\begin{equation}
  f(\theta) = \sum _{L\geq 0} A_L  P_L (\cos \theta), 
\end{equation}
with
\begin{equation}
  A_L = \frac{2L+1}{2} \int _0^\pi d \theta \, \sin \theta \, f( \theta) \, P_L (\cos \theta). 
\end{equation}
It is useful to recall the properties of the Legendre polynomials for which:
\begin{eqnarray}
  && \int_{-1}^1 \left\{(1+x)^{2s}+(1-x)^{2s}\right\} P_{n}(x)\, dx  \nonumber \\
  &&\quad =  \frac{4^{s+1} [\Gamma(2 s+1)]^2}{\Gamma(2s + n + 2) \Gamma (2 s - n +1)} \ \ \ [n = {\rm even}],
\end{eqnarray}
and
  \begin{eqnarray}
    \int_{-1}^1 \left\{(1+x)^{2s}+(1-x)^{2s}\right\} P_{n}(x)\, dx  = 0  \nonumber \\
    \left[n > 2s \ \text{or} \ n = {\rm odd} \right].
\end{eqnarray}
Finally, we find that the isotropized bispectrum can be expressed according to the Legendre-type bispectrum template \eqref{eq:zeta3_cn} with 
\begin{equation}
  \boxed{
    \begin{aligned}
      c_n  & = \frac{4^{s} g_s I_s^0 (2n+1) [\Gamma(2 s+1)]^2}{\Gamma(2s + n + 2) \Gamma (2 s - n +1)} \ \ \ [n = {\rm even}], \\
      c_n &= 0 \ \ \ [n > 2 s \ \text{or} \ n = {\rm odd}].
    \end{aligned}
  }
  \label{eq:cn_theory}
\end{equation}

Now we focus the analysis to the case of partially massless fields. As briefly shown in the introduction, the Higuchi bound implies that higher spin fields perturbations must be suppressed on super-Hubble scales. In case the spinning particles posses particular values of their masses, namely 
\begin{equation}
m^2 = H^2 (s^2-s-2),
\end{equation}
there exists an enhanced symmetry which removes $-1,0,+1$ polarization states. One can show that in this case the spinning particle possesses $\Delta =0$ allowing for the existence of long-lived spin-$s$ perturbations. 
The angle-averaged bispectrum takes the form \cite{Franciolini:2017ktv}:
\begin{equation}
  \Braket{\prod_{i=1}^3 \zeta_{\vec{k}_i} }'_{\text{av}}
  = g_s^{\text{PM}} I^0_s  P_\zeta(k_1)P_\zeta(k_2) \Pi_s+  2~{\rm perm},
\label{eq:bis-pm}
\end{equation}
where 
\begin{equation}
  \Pi_s = \sum_{\lambda_1 \lambda_2\neq 0,\pm1}  d_{s|\lambda_1|}d_{s|\lambda_2|} 
  \left| \epsilon_{\lambda_1, i_1 \cdots i_s}(\vec{k}_1) \epsilon^*_{\lambda_2, i_1 \cdots i_s}(\vec{k}_2) \right|^2\!, 
\end{equation}
The constants $d_{s|\lambda|}$  (with $d_{ss}=1$) represent the mixing of the various helicities into the partial massless field and they arise due to different vacuum expectation values of its corresponding $s - 2$ helicity components.  
For the lowest non-trivial case $s=3$ we find 
\begin{eqnarray}
 \Pi_3 &=& 1 + 15 x^2 +15 x^4
 + x^6 + 18d_{32}(1+5 x^2-5x^4-x^6)\nonumber \\
 &&+9 d_{32}^2(4-15x^2+10x^4+9x^6),
\end{eqnarray}
with $x=\cos \theta_{\hat{k}_1, \hat{k}_2}$, from where one can compute the coefficients $c_n$ entering into Eq.~\eqref{eq:zeta3_cn} to get
\begin{equation}
\boxed{
  \begin{aligned}
    c_{0} &= \frac{16}{7}(2+3 d_{32})^2g_s^{\text{PM}} I^0_s,  \\
    c_2 &= \frac{400}{21} g_s^{\text{PM}} I^0_s , \\
    c_4 &= \frac{72}{77}(2-7 d_{32})^2g_s^{\text{PM}} I^0_s,  \\
    c_6 &= \frac{16}{231}(1-9 d_{32})^2g_s^{\text{PM}} I^0_s,  \\
    c_n &=0 \ \ \ [n \neq 0,2,4,6] .
  \end{aligned}
}
\label{eq:cn_theory_1}
\end{equation}
Similarly for $s=4$ we find
    \begin{eqnarray}
   \Pi_4 &=& 1 + 28 x^2+70 x^4+28 x^6 + x^8 \nonumber \\
   && + 288d_{42}(1+4x^2 - 10 x^4+4 x^6+x^8) \nonumber \\
   && + 16d_{43} (1+14x^2-14 x^6-x^8) \nonumber \\
   && + 576 d_{42}^2(1+6x^2 -11x^4 - 24 x^6+36 x^8)\nonumber \\
   && + 576 d_{43}d_{42}(1-7x^2+15x^4-5 x^6-4x^8) \nonumber \\ 
   && + 512d_{43}^2 , 
    \end{eqnarray}
    and the coefficients $c_n$ of Eq.~\eqref{eq:zeta3_cn} are given in terms of the two parameters $d_{43},d_{42}$ as
    \begin{equation}
\boxed{
 \begin{aligned}
c_{0} =& \frac{256}{315} \Big\{972 d_{42}^2 + 360 d_{42} (1 + d_{43}) \\
&\qquad  +35 (1 + 2 d_{43} + 18 d_{43}^2)\Big\}g_s^{\text{PM}} I^0_s, \\
c_2 =& \frac{512}{693}    \Big\{648 d_{42}^2 + 49 (2 + d_{43})  \\
&\qquad - 72 d_{42} (4 + d_{43})\Big\}  g_s^{\text{PM}} I^0_s , \\
c_4 =& \frac{512}{5005} \Big\{245 + 11763 d_{42}^2 - 735 d_{43} \\
&\qquad + 990 d_{42} (-2 + 3 d_{43})\Big\}g_s^{\text{PM}} I^0_s,  \\
c_6 =& \frac{512}{3465}\Big\{14 + 11664 d_{42}^2 - 119 d_{43}  \\
&\qquad - 198 d_{42} (-4 + 17 d_{43})\Big\}g_s^{\text{PM}} I^0_s,  \\
c_8 =& \frac{128}{6435}(1 + 144 d_{42}) \\
& (1 + 144 d_{42} - 16 d_{43})  g_s^{\text{PM}} I^0_s  ,  \\
c_n =& 0 \ \ \ [n \neq 0,2,4,6,8].
    \end{aligned}
\label{eq:cn_theory_2}
}
\end{equation}
    
\begin{figure*}[t]
  \begin{tabular}{ccc} 
   \begin{minipage}{0.33\hsize}
     \begin{center}
       \includegraphics[width=1\textwidth]{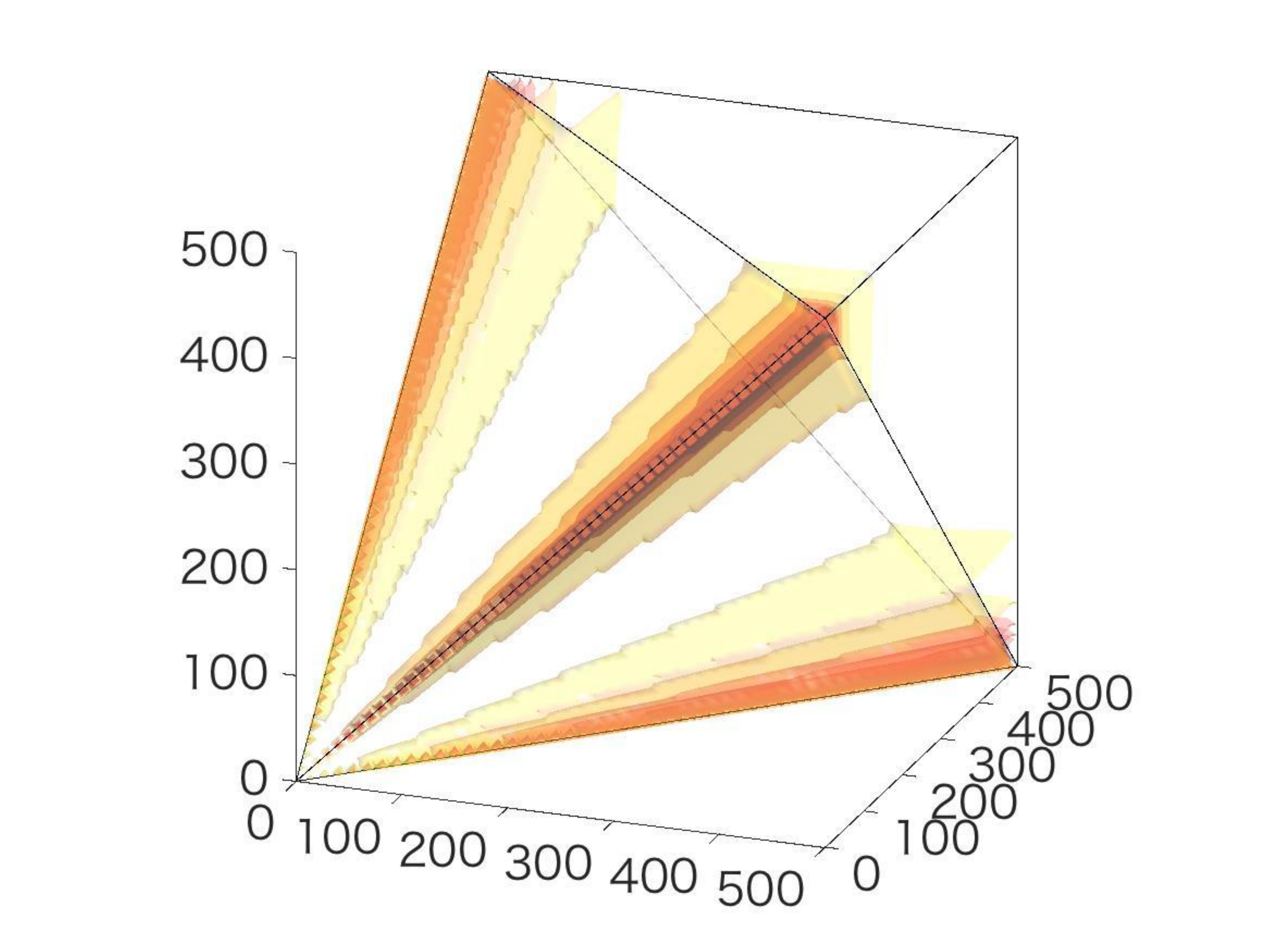}
       ${\cal B}_{k_1 k_2 k_3}^{n = 0}$ or ${\cal B}_{k_1 k_2 k_3}^{f_{\rm NL}^{\rm local}}$
  \end{center}
   \end{minipage}
   \begin{minipage}{0.33\hsize}
     \begin{center}
       \includegraphics[width=1\textwidth]{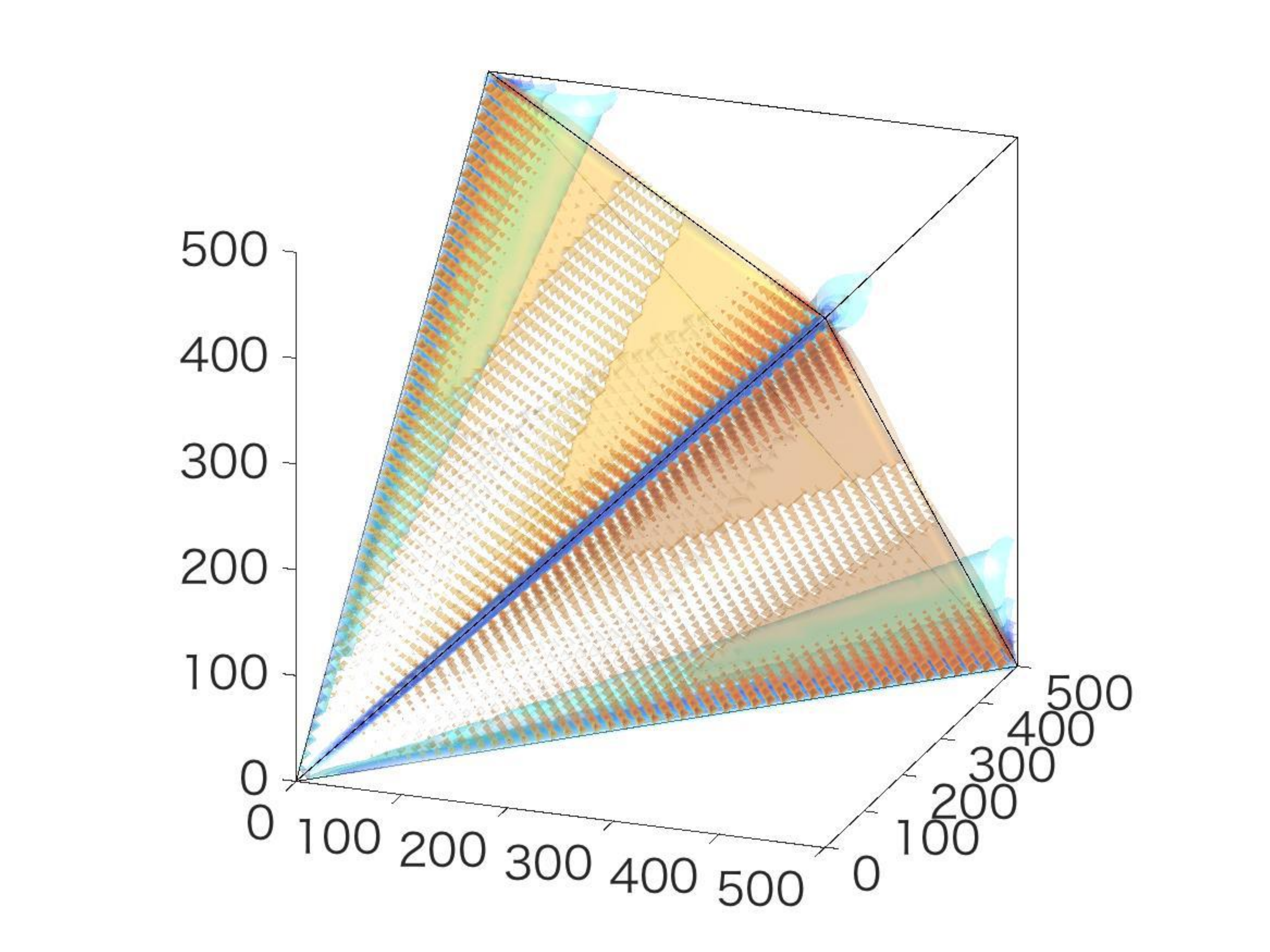}
       ${\cal B}_{k_1 k_2 k_3}^{n = 2}$
  \end{center}
   \end{minipage}
   \begin{minipage}{0.33\hsize}
     \begin{center}
       \includegraphics[width=1\textwidth]{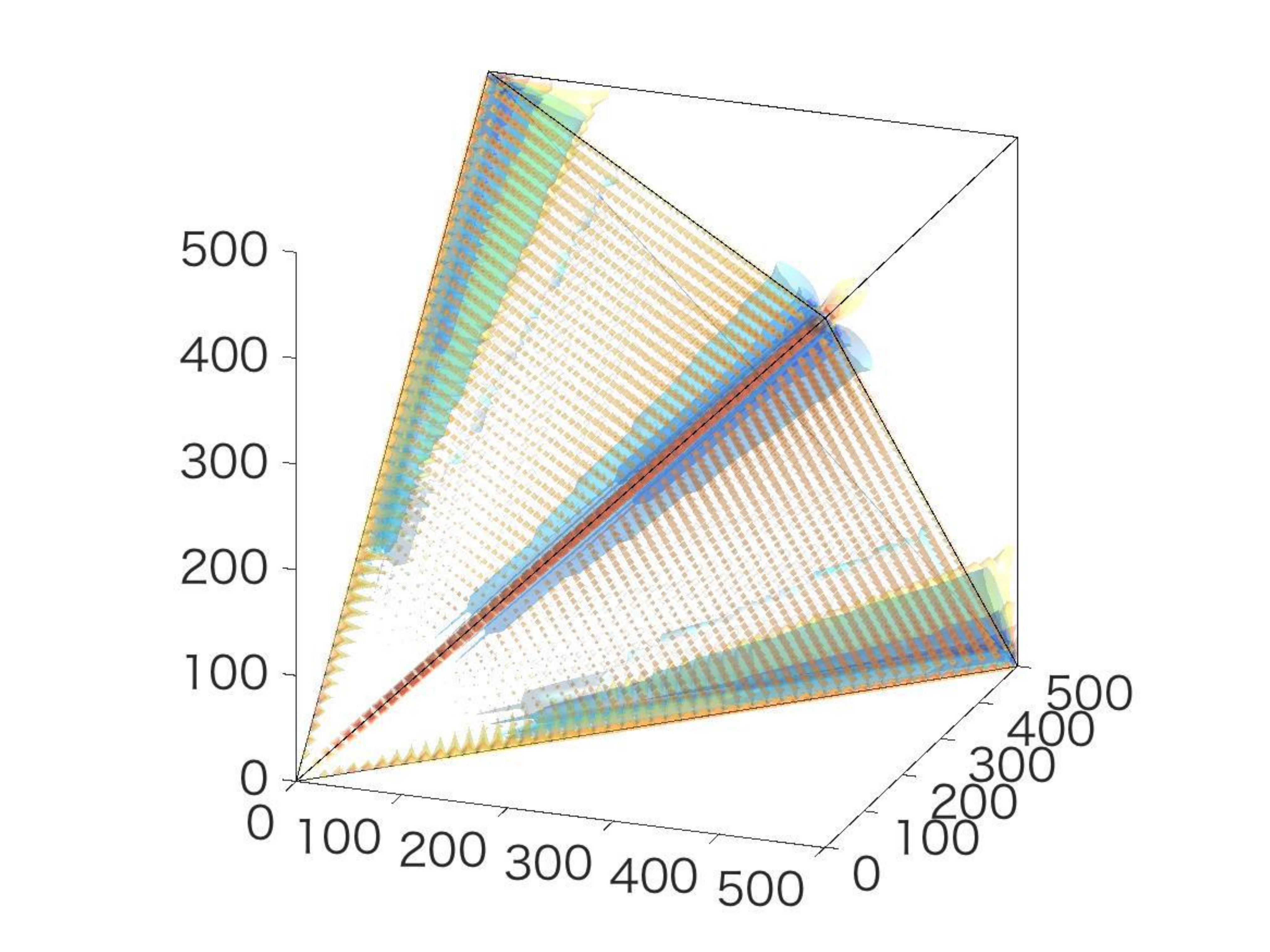}
       ${\cal B}_{k_1 k_2 k_3}^{n = 4}$
     \end{center}
   \end{minipage}
  \end{tabular}
  \\
  \begin{tabular}{ccc} 
   \begin{minipage}{0.33\hsize}
     \begin{center}
       \includegraphics[width=1\textwidth]{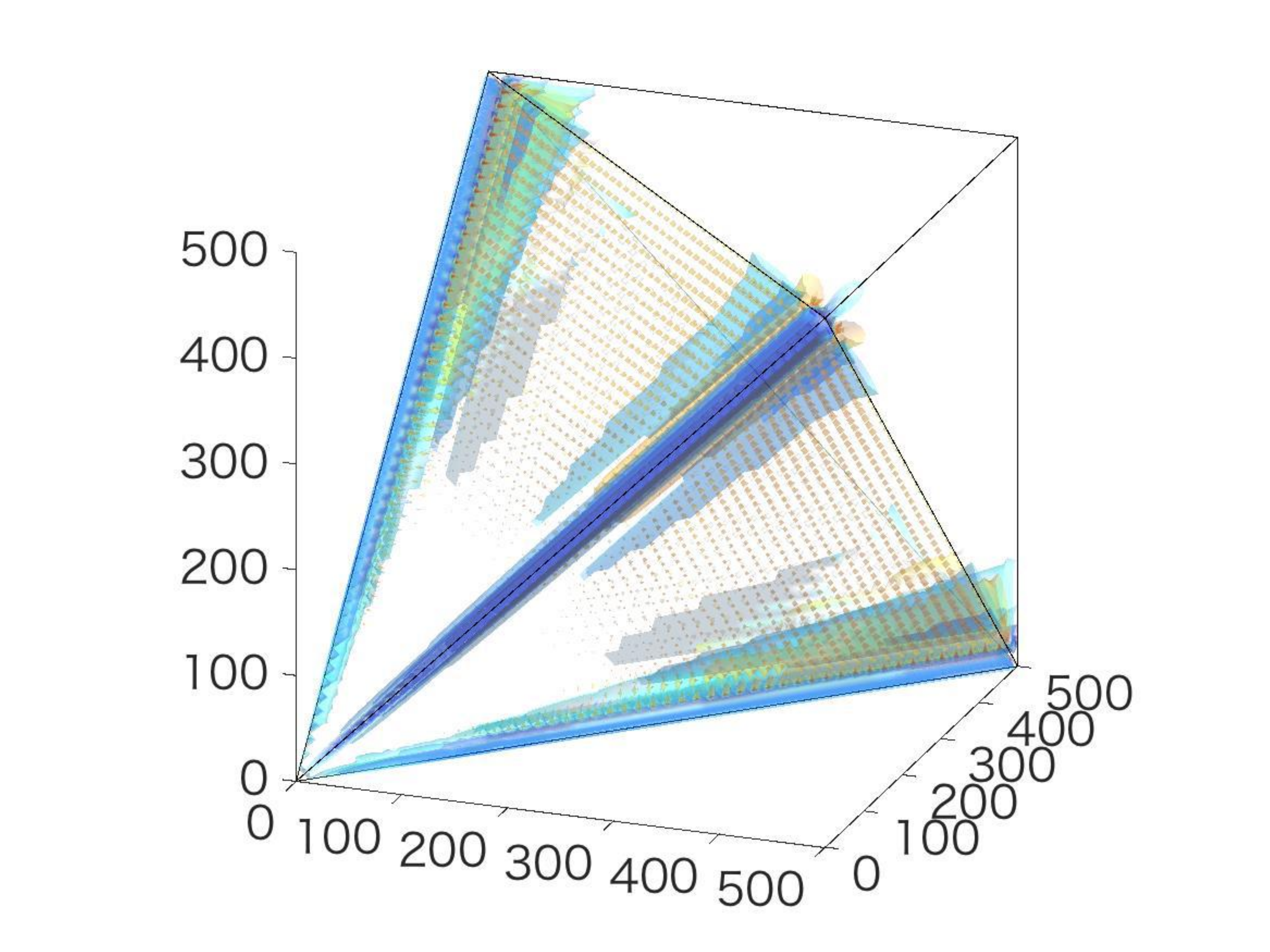}
       ${\cal B}_{k_1 k_2 k_3}^{n = 6}$
     \end{center}
   \end{minipage}
   \begin{minipage}{0.33\hsize}
     \begin{center}
       \includegraphics[width=1\textwidth]{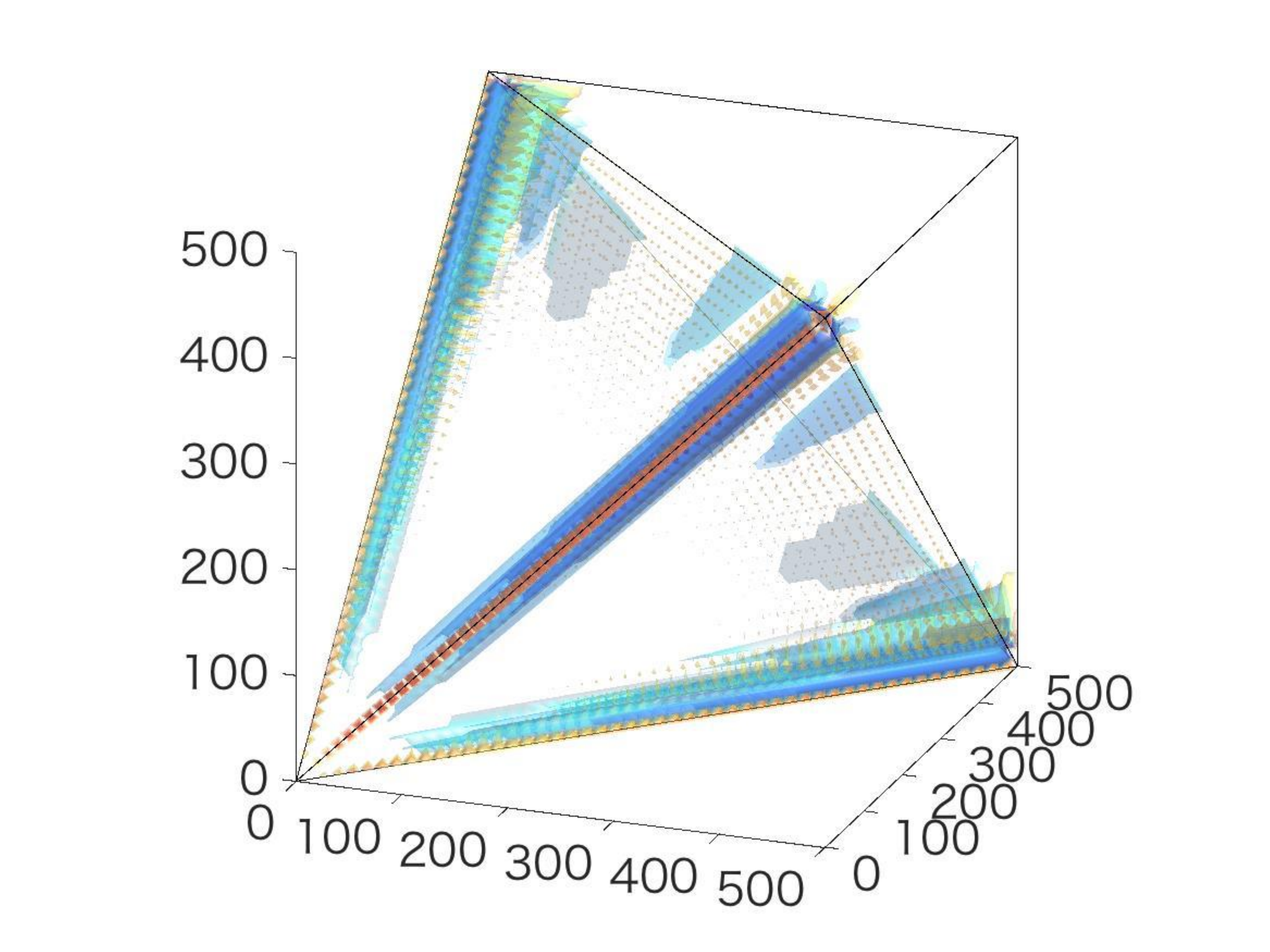}
       ${\cal B}_{k_1 k_2 k_3}^{n = 8}$
  \end{center}
   \end{minipage}
   \begin{minipage}{0.33\hsize}
     \begin{center}
       \includegraphics[width=1\textwidth]{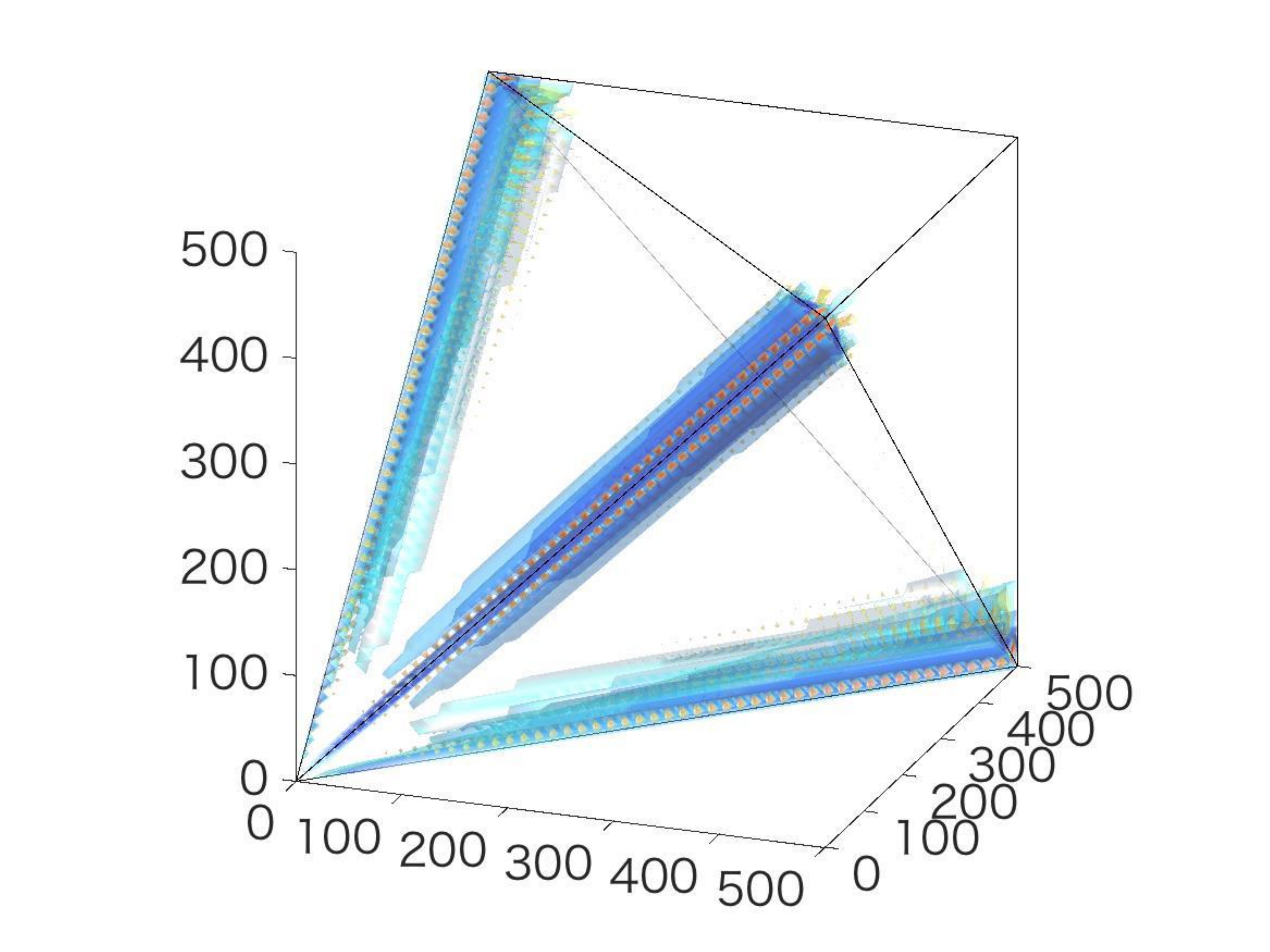}
       ${\cal B}_{k_1 k_2 k_3}^{n = 10}$
     \end{center}
   \end{minipage}
  \end{tabular}
  \caption{Intensity distributions of the primordial curvature bispectra ${\cal B}_{k_1 k_2 k_3}^{n= 0, 2, 4, 6, 8, 10}$ in the $k$-space tetrahedral domain, where the axes correspond to $k_1 r_*$, $k_2 r_*$ and $k_3 r_*$, respectively, with $r_*$ the conformal distance from the last-scattering surface. To highlight the dominant configurations, the bispectrum shapes are rescaled using $k_1^{-2} k_2^{-2} k_3^{-2}$. Dense red (blue) color represents larger positive (negative) signal.}
\label{fig:3D_Bkkk}
\end{figure*}

Figure~\ref{fig:3D_Bkkk} shows the intensity distributions of the curvature bispectrum ${\cal B}_{k_1 k_2 k_3}^n$, given by Eq.~\eqref{eq:zeta3_cn}, in $k$-space tetrahedral domain for $n = 0, 2, 4, 6, 8, 10$. We confirm from this that ${\cal B}_{k_1 k_2 k_3}^{n \geq 2}$ has a strong signal in the squeezed configurations ($k_1 \approx k_2 \gg k_3$, $k_2 \approx k_3 \gg k_1$ and $k_3 \approx k_1 \gg k_2$) and is richer in intensity variation than ${\cal B}_{k_1 k_2 k_3}^{n = 0}$ as it can be inferred from the color code. The latter feature reflects the peculiar angular dependence of Eq.~\eqref{eq:zeta3_cn} for higher $n$. As shown in the next section, these features are directly imprinted in the CMB bispectra.

\section{Imprints in the CMB bispectra}
\label{sec:CMB_bis}

In this section we examine the distinctive signatures of the anisotropic non-Gaussianities due to spinning particles in the CMB bispectra $\Braket{a_{\ell_1 m_1} a_{\ell_2 m_2} a_{\ell_3 m_3}}$, where the CMB multipole coefficients are given by
$a_{\ell m} = \int d^2 \hat{n} \, Y_{\ell m}^*(\hat{n}) \,  T(\hat{n})$, and estimate the detectability of them.

\begin{figure*}[t]
  \begin{tabular}{ccc} 
   \begin{minipage}{0.33\hsize}
     \begin{center}
       \includegraphics[width=1\textwidth]{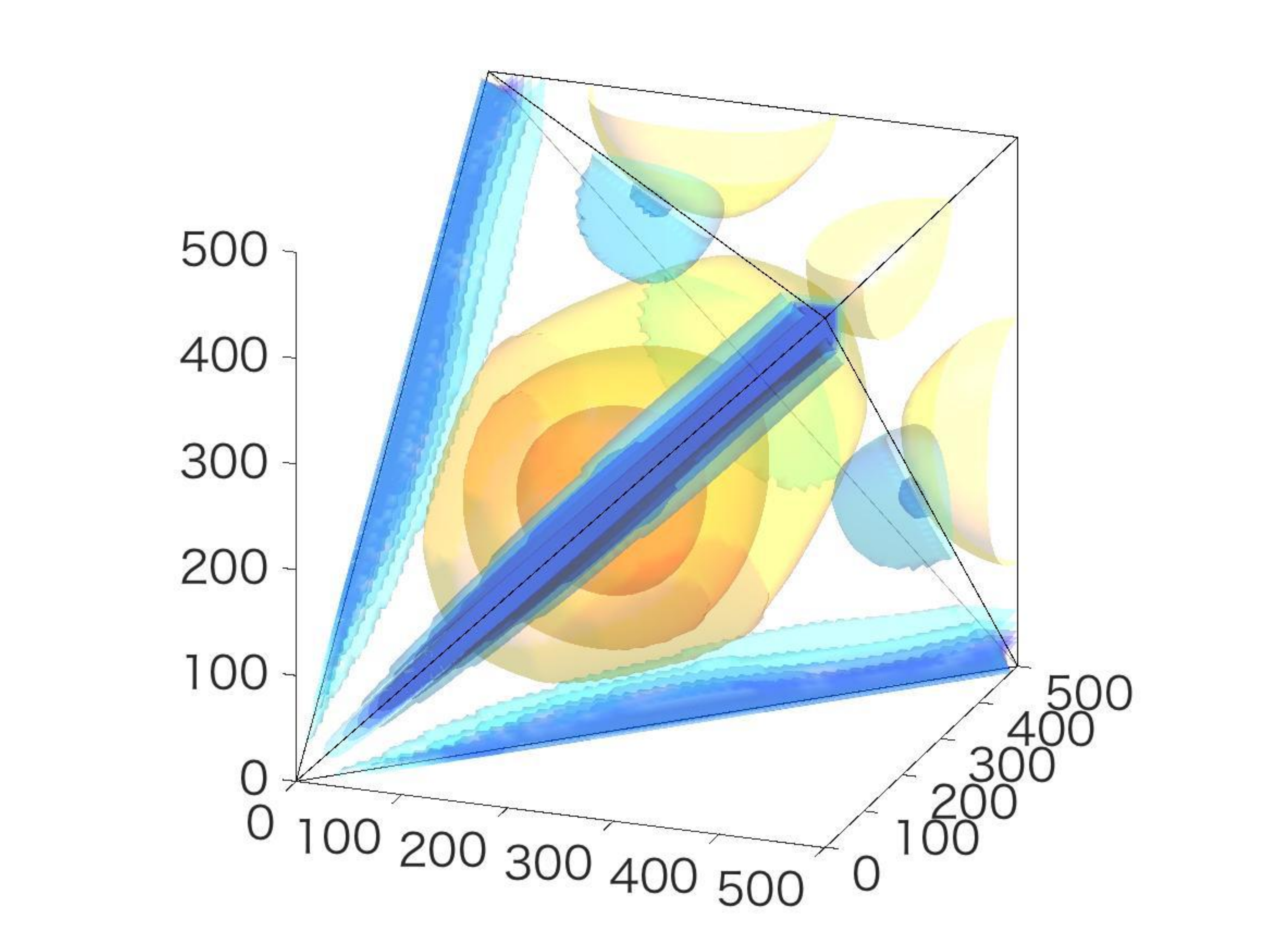}
       $b_{\ell_1 \ell_2 \ell_3}^{n = 0}$ or $b_{\ell_1 \ell_2 \ell_3}^{f_{\rm NL}^{\rm local}}$
  \end{center}
   \end{minipage}
   \begin{minipage}{0.33\hsize}
     \begin{center}
       \includegraphics[width=1\textwidth]{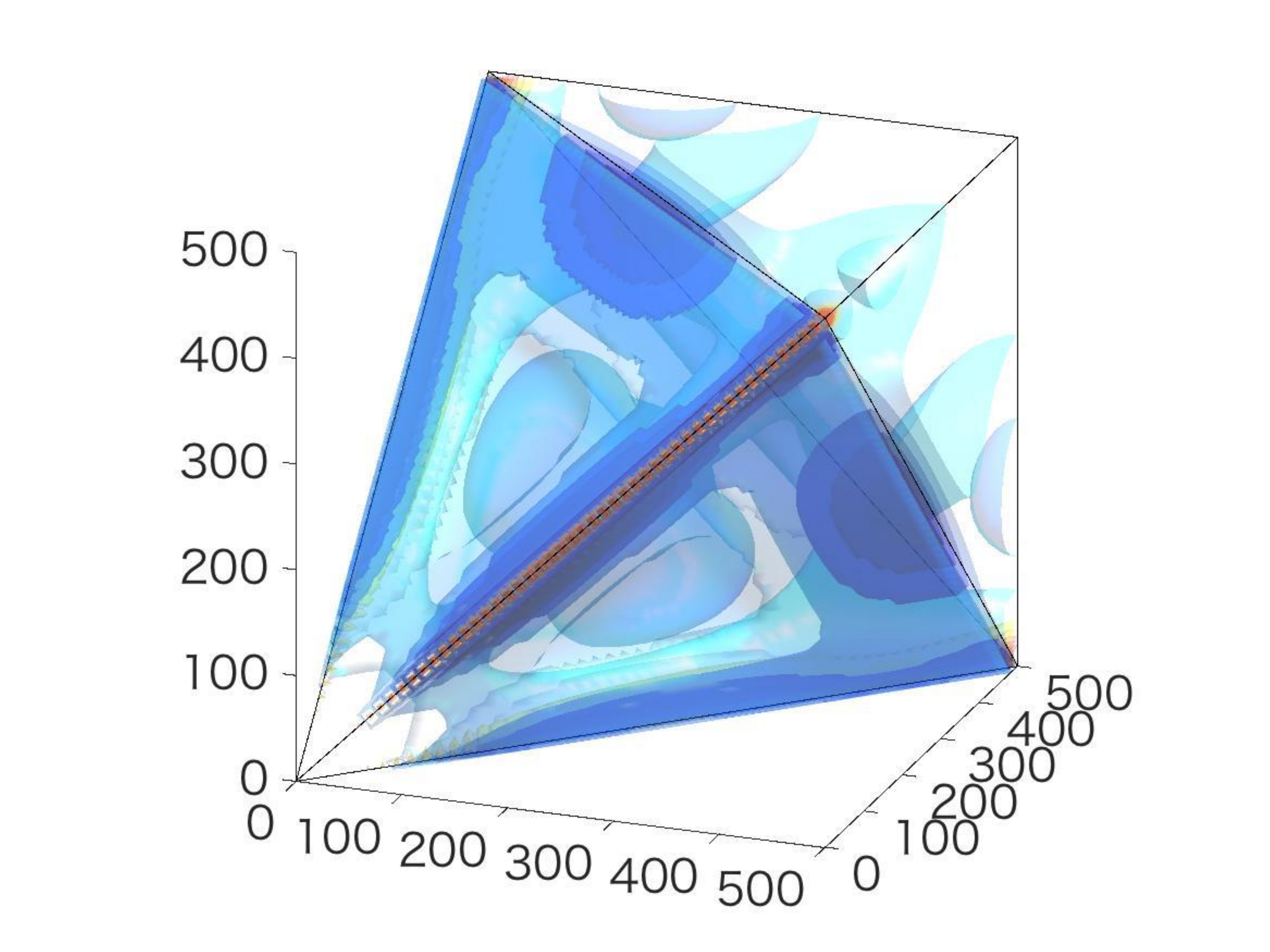}
       $b_{\ell_1 \ell_2 \ell_3}^{n = 2}$
  \end{center}
   \end{minipage}
   \begin{minipage}{0.33\hsize}
     \begin{center}
       \includegraphics[width=1\textwidth]{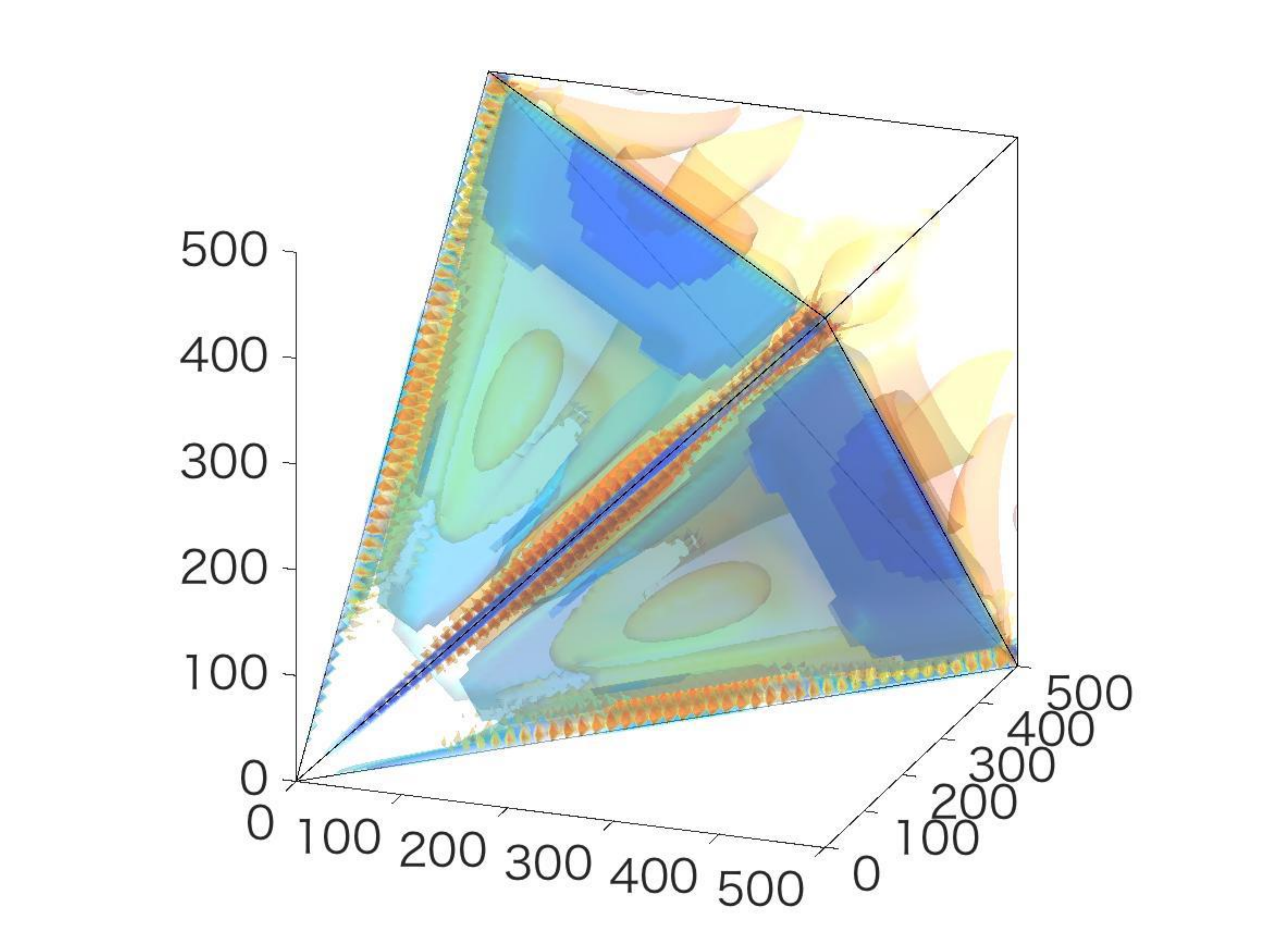}
       $b_{\ell_1 \ell_2 \ell_3}^{n = 4}$
     \end{center}
   \end{minipage}
  \end{tabular}
  \\
  \begin{tabular}{ccc} 
   \begin{minipage}{0.33\hsize}
     \begin{center}
       \includegraphics[width=1\textwidth]{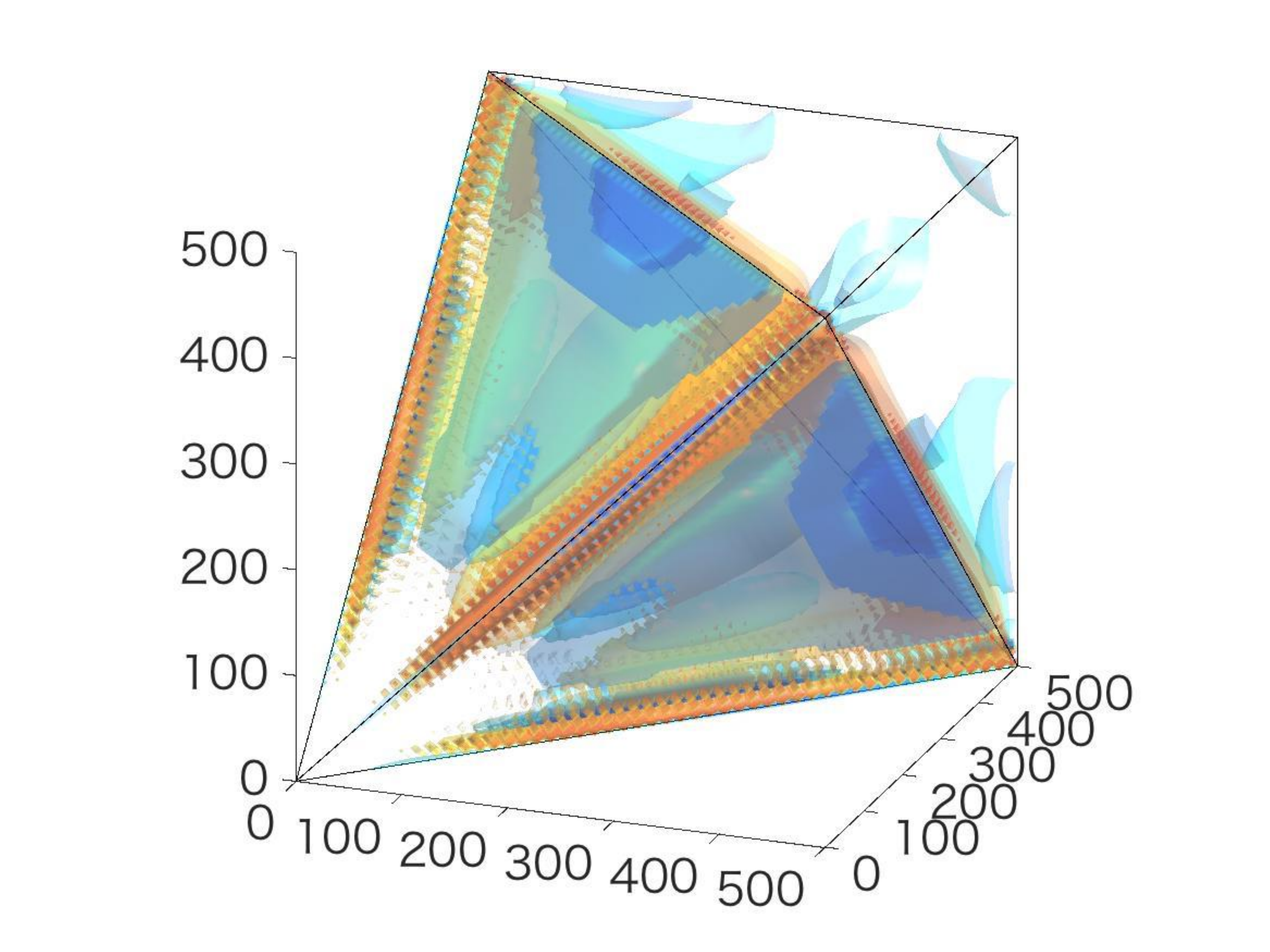}
       $b_{\ell_1 \ell_2 \ell_3}^{n = 6}$
     \end{center}
   \end{minipage}
   \begin{minipage}{0.33\hsize}
     \begin{center}
       \includegraphics[width=1\textwidth]{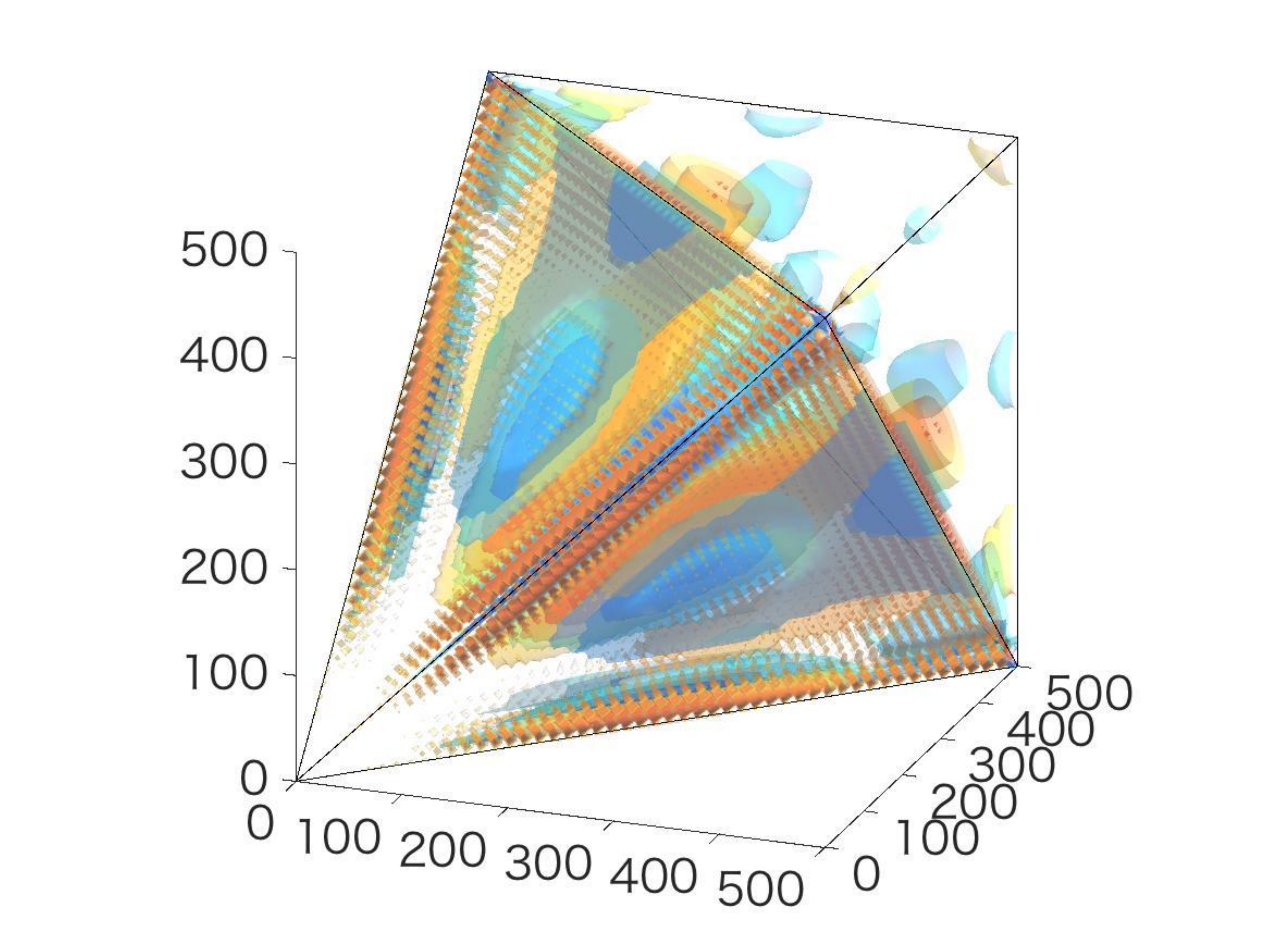}
       $b_{\ell_1 \ell_2 \ell_3}^{n = 8}$
  \end{center}
   \end{minipage}
   \begin{minipage}{0.33\hsize}
     \begin{center}
       \includegraphics[width=1\textwidth]{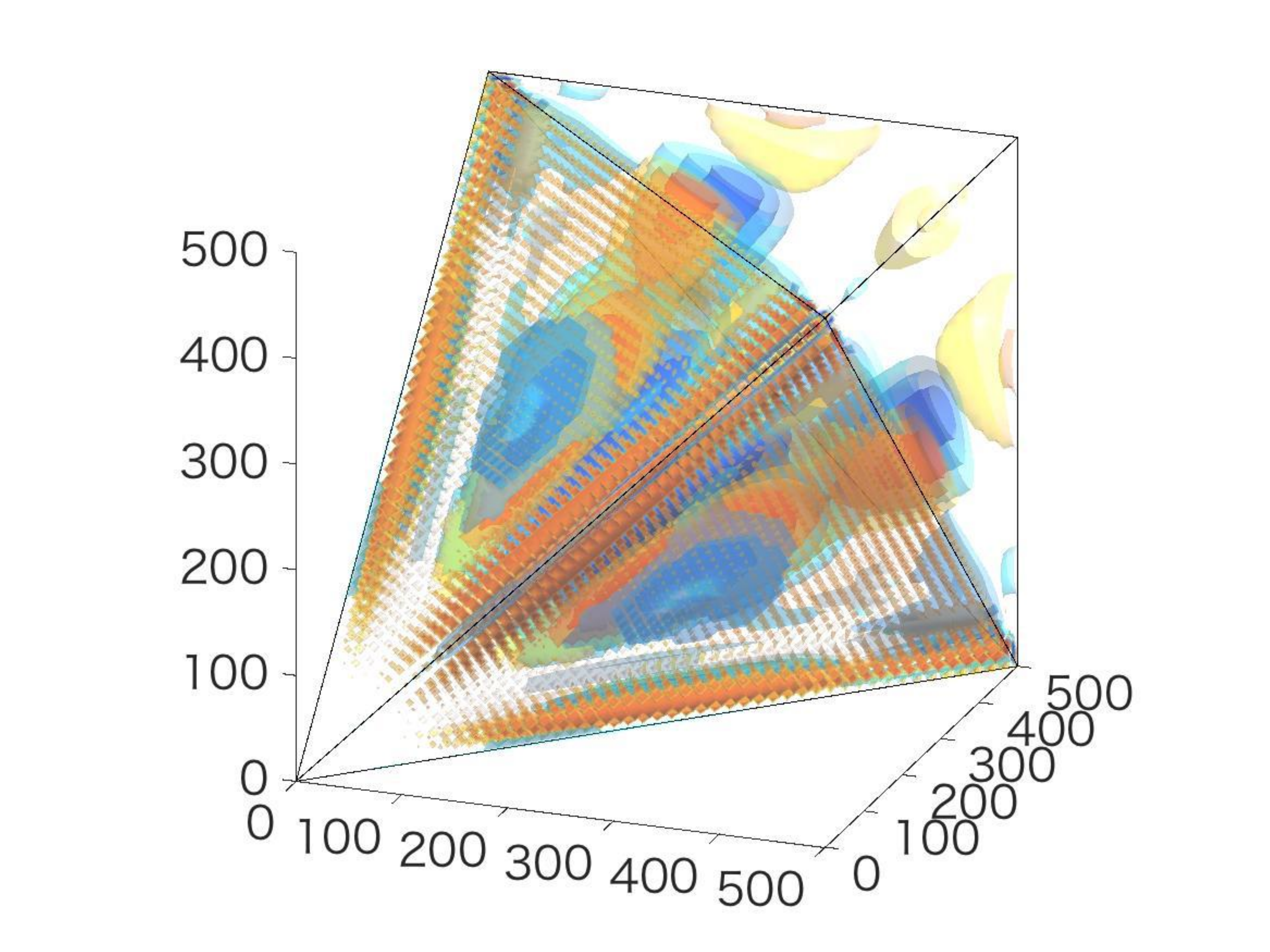}
       $b_{\ell_1 \ell_2 \ell_3}^{n = 10}$
     \end{center}
   \end{minipage}
  \end{tabular}
  \caption{Intensity distributions of the CMB bispectra $b_{\ell_1 \ell_2 \ell_3}^{n= 0, 2, 4, 6, 8, 10}$ in the $\ell$-space tetrahedral domain where the axes correspond to $\ell_1$, $\ell_2$ and $\ell_3$, respectively. For highlighting the dominant configurations, the bispectrum shapes are rescaled using a constant Sachs-Wolfe template \cite{Fergusson:2009nv}. Dense red (blue) color represents larger positive (negative) signal.}
\label{fig:3D_blll}
\end{figure*}

\subsection{Angle-averaged analysis}

As shown in Sec.~\ref{sec:prim}, the curvature bispectrum induced from higher spin fields depends on a peculiar direction $\hat{p}$. This can induce the nonvanishing signal not only inside but also outside the triangular condition $|\ell_2 - \ell_3| \leq \ell_1 \leq \ell_2 + \ell_3$ \cite{Shiraishi:2011ph,Bartolo:2011ee}. The comprehensive analysis including the latter signal is what one should aim for, however, the vastness of the multipole domain that one then has to consider makes it unfeasible. We therefore focus on the former signal alone by using the usual angle-averaged bispectrum form:
\begin{equation}
  B_{\ell_1 \ell_2 \ell_3} = \sum_{m_1 m_2 m_3}
\left(
  \begin{array}{ccc}
  \ell_1 & \ell_2 & \ell_3 \\
  m_1 & m_2 & m_3 
  \end{array}
  \right) 
  \Braket{\prod_{i = 1}^3 a_{\ell_i m_i} }.
\end{equation}
This is mathematically equivalent to the CMB bispectrum computed from the ``isotropized'' curvature bispectrum given by Eqs.~\eqref{eq:zeta3_cn}, \eqref{eq:cn_theory}, \eqref{eq:cn_theory_1} and \eqref{eq:cn_theory_2}. In the following, we compute this CMB bispectrum following the methodology in Refs.~\cite{Shiraishi:2010sm,Shiraishi:2010kd,Shiraishi:2013vja}.

The CMB coefficients computed from the primordial curvature perturbation are expressed as
\begin{equation}
  a_{\ell m} =
4\pi (-i)^{\ell} \int \frac{d^3 \vec{k}}{(2\pi)^{3}}
{\cal T}_{\ell}(k) \zeta_{\vec{k}}  Y_{\ell m}^*(\hat{k}).
\end{equation}
The angle-averaged CMB bispectrum sourced from Eq.~\eqref{eq:zeta3_cn} is therefore given by $B_{\ell_1 \ell_2 \ell_3} = \sum_n c_n B_{\ell_1 \ell_2 \ell_3}^n$ with 
\begin{eqnarray}
 && B_{\ell_1 \ell_2 \ell_3}^n = \sum_{m_1 m_2 m_3}
\left(
  \begin{array}{ccc}
  \ell_1 & \ell_2 & \ell_3 \\
  m_1 & m_2 & m_3 
  \end{array}
  \right) \nonumber \\
  &&\qquad \left[ \prod_{j=1}^3 \frac{(-i)^{\ell_j}}{\pi} \int d^3 \vec{k}_j  {\cal T}_{\ell_j}(k_j) Y_{\ell_j m_j}^*(\hat{k}_j) \right]
  \delta^{(3)}\left(\sum_{i=1}^3 \vec{k}_i \right)
  \nonumber \\ 
  &&\qquad  
  \left[ P_n(\hat{k}_1 \cdot \hat{k}_2) P_\zeta(k_1)P_\zeta(k_2) + 2~{\rm perm} \right]  ,
\end{eqnarray}
where ${\cal T}_{\ell}(k)$ is the scalar-mode temperature transfer function. It is then convenient to decompose the $\hat{k}$-dependent parts into the spherical harmonics by use of the identities: 
\begin{equation}
 P_n(\hat{k}_1 \cdot \hat{k}_2) 
 = \frac{4\pi}{2n+1} \sum_{m} Y_{n m}^* (\hat{k}_1) Y_{n m} (\hat{k}_2), 
\end{equation}
and 
\begin{eqnarray}
&& \delta^{(3)}\left( \sum_{i=1}^3 \vec{k}_i \right) 
= 8 \int_0^\infty r^2 dr \nonumber \\ 
&&\qquad  \left[ \prod_{i=1}^3 \sum_{L_i M_i} j_{L_i}(k_i r) Y_{L_i M_i}^*(\hat{k}_i) \right] \nonumber \\ 
&&\qquad (-1)^{\frac{L_1 + L_2 +L_3}{2}}
h_{L_1 L_2 L_3}
\left(
  \begin{array}{ccc}
  L_1 & L_2 & L_3 \\
  M_1 & M_2 & M_3 
  \end{array}
 \right),
\end{eqnarray}
where
\begin{equation}
  h_{\ell_1 \ell_2 \ell_3}
\equiv \sqrt{\frac{(2 \ell_1 + 1)(2\ell_2 + 1)(2 \ell_3 + 1)}{4 \pi}}
\left(
  \begin{array}{ccc}
  \ell_1 & \ell_2 & \ell_3 \\
  0 & 0 & 0
  \end{array}
  \right).
  \end{equation}
After performing the angular $\hat{k}$ integrals of the products of the spherical harmonics and the additions of the induced angular momenta by use of the identities:
\begin{eqnarray}
 \int d^2 \hat{k}  \prod_{i = 1}^2 Y_{\ell_i m_i}^*(\hat{k}) &=& (-1)^{m_1}\delta_{\ell_1, \ell_2} \delta_{m_1, -m_2} ~, \\
 \int d^2 \hat{k} \prod_{i = 1}^3  Y_{\ell_i m_i}^*(\hat{k}) &=& h_{\ell_1\ell_2 \ell_3}
  \left(
  \begin{array}{ccc}
  \ell_1 & \ell_2 & \ell_3 \\
  m_1 & m_2 & m_3 
  \end{array}
  \right) ,
  \end{eqnarray}
and
\begin{eqnarray}
&& \sum_{m_4 m_5 m_6} (-1)^{\sum_{i=4}^6( \ell_i - m_i) }
\left(
\begin{array}{ccc}
  \ell_5 & \ell_1 & \ell_6 \\
  m_5 & -m_1 & -m_6 
 \end{array}
 \right) \nonumber \\
&&\quad \left(
\begin{array}{ccc}
  \ell_6 & \ell_2 & \ell_4 \\
  m_6 & -m_2 & -m_4 
\end{array}
  \right)
\left(
 \begin{array}{ccc}
  \ell_4 & \ell_3 & \ell_5 \\
  m_4 & -m_3 & -m_5 
\end{array}
 \right) \nonumber \\
 &&\qquad
 = \left(
  \begin{array}{ccc}
  \ell_1 & \ell_2 & \ell_3 \\
  m_1 & m_2 & m_3 
 \end{array}
 \right) 
\left\{
 \begin{array}{ccc}
  \ell_1 & \ell_2 & \ell_3 \\
  \ell_4 & \ell_5 & \ell_6 
 \end{array}
 \right\}, 
\end{eqnarray}
we obtain the simple expression found in Ref.~\cite{Shiraishi:2013vja}, reading
\begin{eqnarray}
 B_{\ell_1 \ell_2 \ell_3}^{n} &=&  
\int_0^\infty r^2 dr 
\left[ \prod_{i=1}^3 \sum_{L_i} 
 (-1)^{\frac{\ell_i + L_i}{2}}  \right]   h_{L_1 L_2 L_3} \nonumber \\ 
&& \beta_{\ell_1 L_1}(r) \beta_{\ell_2 L_2}(r) \alpha_{\ell_3}(r)   
\frac{4\pi}{2n+1}  
h_{\ell_1 L_1 n}
 h_{\ell_2 L_2 n}  \nonumber \\ 
&& 
(-1)^{\ell_2 + L_1} \delta_{L_3, \ell_3}
\left\{
  \begin{array}{ccc}
  \ell_1 & \ell_2 & \ell_3 \\
  L_2 & L_1 & n 
  \end{array}
 \right\}
+ 2~{\rm perm}, \label{eq:CMB_bis_all}
\end{eqnarray}
where 
\begin{eqnarray}
\alpha_{\ell}(r) &\equiv& \frac{2}{\pi} 
\int_0^\infty k^2 dk {\cal T}_{\ell}(k) j_\ell(k r), \label{eq:alpha} \\ 
\beta_{\ell L}(r) &\equiv& \frac{2}{\pi} 
\int_0^\infty k^2 dk P_{\zeta}(k) {\cal T}_{\ell}(k) j_L(k r). \label{eq:beta}
\end{eqnarray}

Figure~\ref{fig:3D_blll} describes the intensity distributions of the reduced bispectra $b_{\ell_1 \ell_2 \ell_3}^n \equiv B_{\ell_1 \ell_2 \ell_3}^n / h_{\ell_1 \ell_2 \ell_3}$ in $\ell$-space tetrahedral domain for $n= 0, 2, 4, 6, 8, 10$. The signal is peaked around the squeezed configurations ($\ell_1 \approx \ell_2 \gg \ell_3$, $\ell_2 \approx \ell_3 \gg \ell_1$ and $\ell_3 \approx \ell_1 \gg \ell_2$). This is due to the squeezed-limit magnification of the curvature bispectra confirmed from Fig.~\ref{fig:3D_Bkkk}. As seen in the next subsection, this fact enhances the detectability of $c_n$. Moreover, as expected from Fig.~\ref{fig:3D_Bkkk}, the rich intensity variation of $b_{\ell_1 \ell_2 \ell_3}^{n \geq 2}$ around the squeezed configurations is also confirmed. Actually, the correlation between $b_{\ell_1 \ell_2 \ell_3}^{n \geq 2}$ and $b_{\ell_1 \ell_2 \ell_3}^{n = 0}$ is $\lesssim 10 \%$, indicating the difference in shape. Note that the  intensity variations in  the configurations (except the squeezed ones) observed here (but unobserved in Fig.~\ref{fig:3D_Bkkk}) are mainly caused by the acoustic oscillations at the recombination epoch. Such configurations though do not give a major contribute  to the signal-to-noise ratio.

Because of the existence of the $r$ integrals, especially for higher $n$, Eq.~\eqref{eq:CMB_bis_all} requires unfeasibly massive computational cost. For cost savings, we also use the Sachs-Wolfe (SW) approximation: ${\cal T}_{\ell}(k) \approx - \frac{1}{5} j_\ell (k r_*)$ with $r_*$ the conformal distance from the last-scattering surface, leading to \cite{Shiraishi:2013vja}
\begin{eqnarray}
 B_{\ell_1 \ell_2 \ell_3}^{n, \rm SW} &=&  
- \frac{1}{5} 
\left[ \prod_{i=1}^3 \sum_{L_i} 
 (-1)^{\frac{\ell_i + L_i}{2}}  \right] 
h_{L_1 L_2 L_3} \beta_{\ell_1 L_1}^{\rm SW} \beta_{\ell_2 L_2}^{\rm SW} \nonumber \\ 
&& 
\frac{4\pi}{2n+1}  
h_{\ell_1 L_1 n} h_{\ell_2 L_2 n} (-1)^{\ell_2 + L_1} \delta_{L_3, \ell_3} \nonumber \\ 
&&  
\left\{
  \begin{array}{ccc}
  \ell_1 & \ell_2 & \ell_3 \\
  L_2 & L_1 & n 
  \end{array}
 \right\}
+ 2~{\rm perm} , \label{eq:CMB_bis_SW}
\end{eqnarray}
with 
\begin{eqnarray}
  \beta_{\ell L}^{\rm SW} \equiv - \frac{\pi^2}{10} {\cal P} 
\frac{\Gamma\left(\frac{\ell + L}{2}\right)}{\Gamma\left(\frac{\ell - L + 3}{2}\right) \Gamma\left(\frac{-\ell + L + 3}{2}\right) 
\Gamma\left(\frac{\ell + L + 4}{2}\right)}.
\end{eqnarray}
To derive this, $P_\zeta(k) = 2\pi^2 {\cal P} k^{-3}$ has been assumed.


\begin{figure}[t]
  \begin{tabular}{c}
    \begin{minipage}{1.\hsize}
  \begin{center}
    \includegraphics[height = 6.8cm]{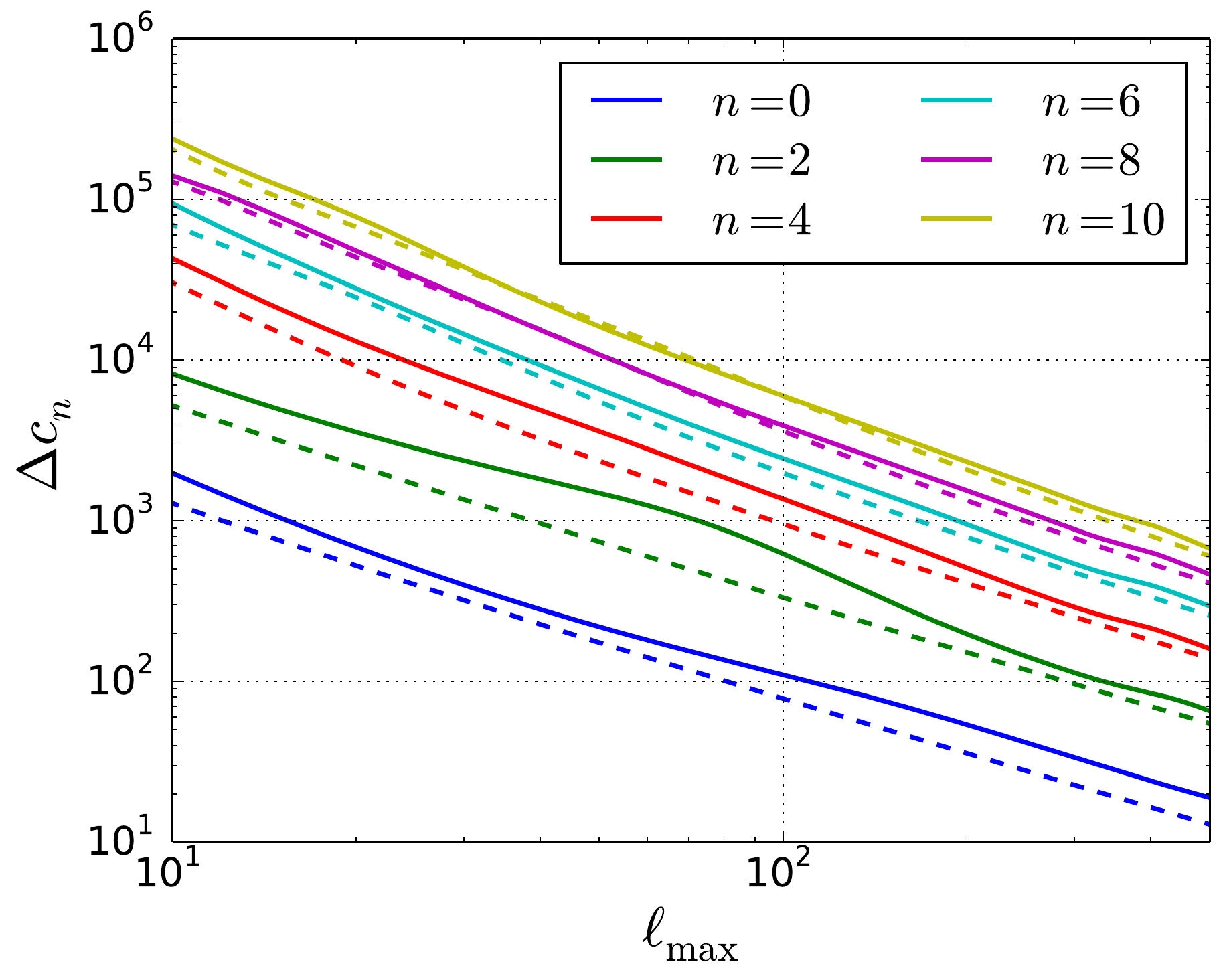}
  \end{center}
\end{minipage}
\end{tabular}
  \caption{Expected $1\sigma$ errors on $c_{0, 2, 4, 6, 8, 10}$ as a function of $\ell_{\rm max}$. The solid and dashed lines are computed using the exact expression~\eqref{eq:CMB_bis_all} and the SW formula~\eqref{eq:CMB_bis_SW}, respectively. The results for $n = 2$ are consistent with those obtained in Ref.~\cite{Shiraishi:2013vja}.}\label{fig:error_cn_vs_lmax} 
\end{figure}


\subsection{Sensitivity to $c_n$}

For the forecast analysis, we assume the measurement of very weak NG signal. The covariance of the bispectrum can then be approximately expressed with the product of the angular power spectra $C_\ell$; thus, the Fisher matrix for $c_n$ reads
\begin{eqnarray}
  F_{n} = \sum_{\ell_1, \ell_2, \ell_3 = 2}^{\ell_{\rm max}} \frac{(B_{\ell_1 \ell_2 \ell_3}^n)^2}{6 C_{\ell_1} C_{\ell_2} C_{\ell_3}} . \label{eq:fish}
\end{eqnarray}
We here assume a full-sky cosmic-variance-limited-level (CVL-level) measurement, so $C_\ell$ does not contain any specific instrumental information. The expected $1\sigma$ error on $c_n$ is then computed according to $\Delta c_n = 1 / \sqrt{F_n}$.

\begin{figure}[t]
  \begin{tabular}{c}
    \begin{minipage}{1.\hsize}
  \begin{center}
    \includegraphics[height = 6.8cm]{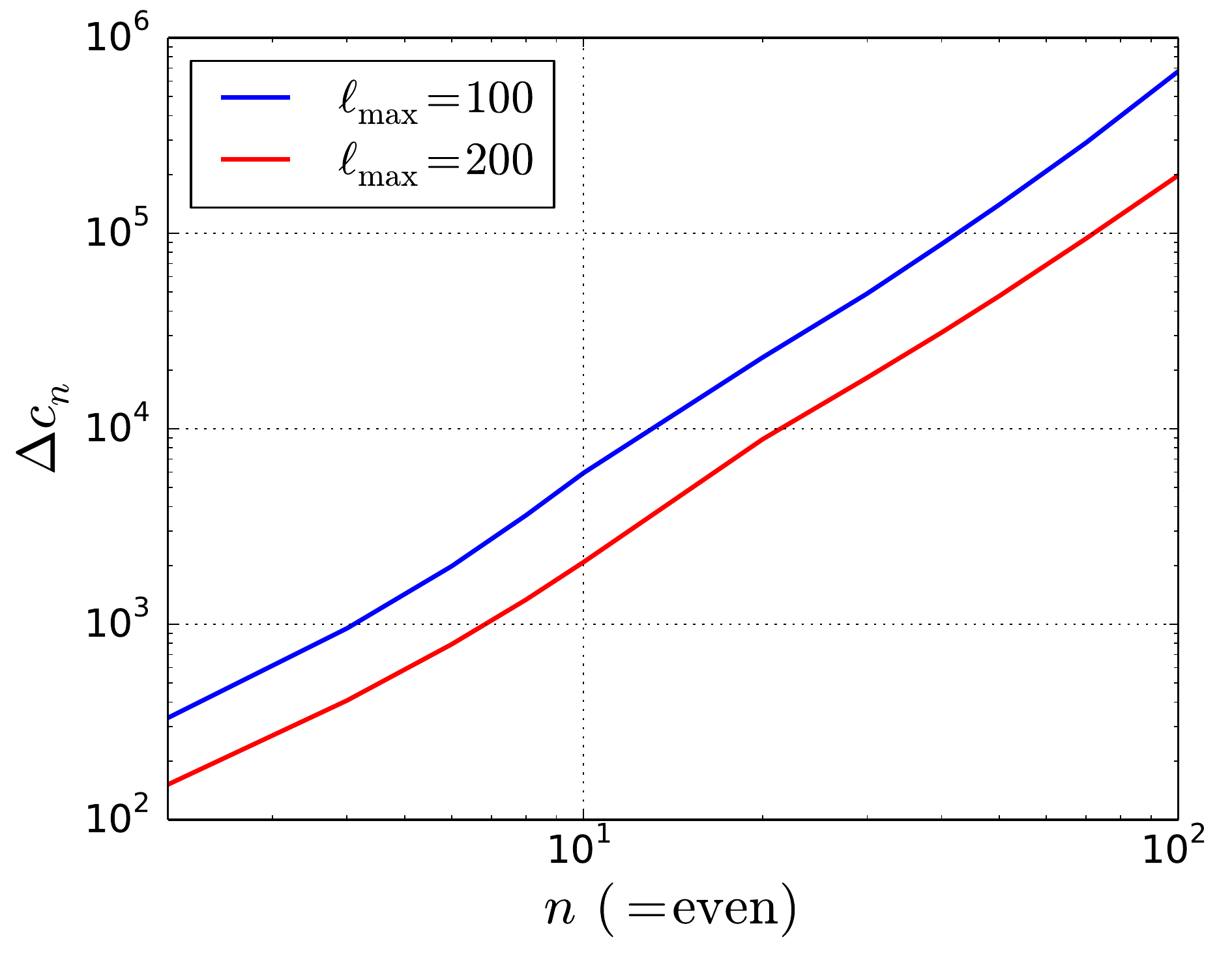}
  \end{center}
\end{minipage}
\end{tabular}
  \caption{Expected $1\sigma$ errors on $c_{n = \rm even}$ at $\ell_{\rm max} = 100$ and $200$ as a function of $n$. These lines are estimated using the SW formula~\eqref{eq:CMB_bis_SW}.}\label{fig:error_cn_vs_n} 
\end{figure}

Figure~\ref{fig:error_cn_vs_lmax} shows $\Delta c_{0, 2, 4, 6, 8, 10}$ as a function of $\ell_{\rm max}$. It is confirmed that, even for higher $n$, $\Delta c_n$ roughly follows the scaling relation expected in the squeezed-type NG case, i.e., $\Delta c_n \propto \ell_{\rm max}^{-1} (\ln \ell_{\rm max})^{-1/2}$ \cite{Komatsu:2001rj,Shiraishi:2010kd,Shiraishi:2013vja}. The consistency between the solid line and the dashed one for each $n$ that is observed here shows the validity of the use of the SW approximation. This figure also indicates that, at any fixed $\ell_{\rm max}$, $\Delta c_n$ is an increasing function of $n$. This is confirmed from Fig.~\ref{fig:error_cn_vs_n} that draws $\Delta c_n(\ell_{\rm max} = 100)$ and $\Delta c_n(\ell_{\rm max} = 200)$ as a function of $n$ computed from the SW formula~\eqref{eq:CMB_bis_SW}. We here find a scaling relation as $\Delta c_n \propto n^2$.
From the above estimates, we can expect that $\Delta c_n$ behaves like 
\begin{equation}
  \boxed{
    \begin{aligned}
  \Delta c_n \sim 2 \, n^2 \, \left(\frac{2000}{\ell_{\rm max}}\right) \left(\ln \frac{\ell_{\rm max}}{2000}\right)^{-1/2} \\
  [n \geq 2 \ \text{and} \ n = {\rm even}] .
  \end{aligned} 
} \label{eq:error_cn}
\end{equation}
This result indicates, for instance, that a $c_2$ as large as $\sim 10$ is measurable  using the CMB bispectrum. From {\it Planck} data a null signal has been detected with a comparable accuracy \cite{Ade:2013ydc,Ade:2015ava}. This may be further improved if the small-scale data of density fluctuations becomes reachable by future surveys of galaxies \cite{Byun:2014cea,Raccanelli:2015oma,Schmidt:2015xka,Chisari:2016xki,MoradinezhadDizgah:2018ssw} and the 21-cm anisotropies \cite{Munoz:2015eqa}.%
\footnote{For instance, Ref.~\cite{MoradinezhadDizgah:2018ssw} recently  reported encouraging results on the  detectability of signals in  the galaxy bispectrum measurements of higher spin fields up to $s=4$ and within the effective field theory approach.}


\section{Conclusions}\label{sec:con}

Higher spin fields may have been present during a period of primordial inflation. If they also happened to be quantum-mechanically excited with long-living 
perturbations on scales larger than the Hubble radius, they may have left an imprint on the curvature perturbations and consequently on the CMB anisotropy and LSS. In this paper we have extended the previous analysis on the impact of higher spin fields on the power spectra of CMB and LSS \cite{Bartolo:2017sbu} by studying the statistical anisotropy imprinted in the CMB bispectrum. Our findings are summarized by the forecast in Eq.~\eqref{eq:error_cn}  where we stress that the errors on the coefficients  $c_{n \geq 4}$ have been obtained for the first time. It would be also nice to investigate the impact of higher spin fields within the so-called Vasiliev construction  \cite{Vasiliev:1990en}  which comes with an infinite tower of massless higher spin states.


\acknowledgements

We thank James R. Fergusson for helping to draw the  3D representations of the bispectra. M.\,S. is supported by a JSPS Grant-in-Aid for Research Activity Start-up Grant No.~17H07319. M.\,S. also acknowledges the Center for Computational Astrophysics, National Astronomical Observatory of Japan, for providing the computing resources of Cray XC30. A.\,R. is supported by the Swiss National Science Foundation (SNSF), project {\sl Investigating the Nature of Dark Matter}, Project No.~200020-159223. 

\appendix

\begin{widetext}
  
\section{Higher spin polarization tensors and projectors}\label{app:1}

In this appendix we review the higher spin projector tensors which were explicitly constructed in Ref.~\cite{Franciolini:2017ktv}. The polarization tensors of higher spin fields ($\lambda$ indices the various polarizations) are
\begin{equation}
  \begin{aligned}
    \epsilon_\lambda^{i_1\cdots i_s}(\vec{k})=&\sum_{\lambda_1,\cdots,\lambda_s=\pm 1}\delta_{\lambda_1+\cdots+\lambda_s,\lambda}
    \sqrt{\frac{2^s(s+\lambda)!(s-\lambda)!}{(2s)!\prod_{i=1}^s(1+\lambda_i)!(1-\lambda_i)!}}\prod_{j=1}^s\epsilon_{\lambda_j}^{i_j}(\vec{k}), \\
    \epsilon_\lambda^{*i_1\cdots i_s}(\vec{k})=&\sum_{\lambda_1,\cdots,\lambda_s=\pm 1}\delta_{\lambda_1+\cdots+\lambda_s,\lambda}
    \sqrt{\frac{2^s(s+\lambda)!(s-\lambda)!}{(2s)!\prod_{i=1}^s(1+\lambda_i)!(1-\lambda_i)!}}\prod_{j=1}^s \epsilon_{\lambda_j}^{*i_j}(\vec{k}), \label{eq:app.1}
\end{aligned}
\end{equation}
which are written in terms of the polarization tensors for a spin-1 field $\epsilon_{\lambda}^{i}(\vec{k})$. Clearly, the polarization tensors $\epsilon^\lambda_{i_1\cdots i_s} (\vec{k})$ are totally symmetric, traceless and satisfy $k^{i_1}\epsilon^\lambda_{i_1\cdots i_s}(\vec{k})=0$. The projector tensor in $d$ dimensions can be defined as:
\begin{equation}
\Pi^{ i_1 \cdots i_s j_1 \cdots j_s} (\vec{k})
\equiv \sum_{\lambda} \epsilon_\lambda ^{i_1 \cdots i_s} (\vec{k}) \epsilon^{* j_1 \cdots j_s}_{\lambda } (\vec{k}). \label{eq:proj}
\end{equation}
It was shown that it can be expressed in terms of the spin-1 projector tensor $\Pi^{ij} (\vec{k}) \equiv \delta^{ij} - {k^i k^j}/{k^2}$:
\begin{equation}
    \Pi^{i_1 \cdots i_s j_1 \cdots j_s} (\vec{k})=
    \left( \frac{1}{s!}\right) \sum_{P{(i)}P{(j)}}
    \left[
      \sum_{r=0  }^{r\leq\frac{s}{2}} 
C(s,r) \Pi^{i_1 i_2}\Pi^{j_1 j_2} \cdots \Pi^{i_{2r-1} i_{2r}}\Pi^{j_{2r-1} j_{2r}} \prod_{n=2r+1}^{s} \Pi^{i_n j_n} \right],
  \label{eq:bc}
\end{equation}
where the symbol $P(l)$ identify permutations of the index $l$. The coefficients determining Eq.~\eqref{eq:bc} are: 
\begin{equation}
  \begin{aligned}
    &C(s,r) =- \left\{\frac{ C(s,r-1)A(s,r-1)A(s-2,r-1) [s-2(r-1)]!}{A(s,r) (s-2r)!\left[A(s,r)-A(s-2,r)+  (d-2) A(s-2,r-1)\right]} \right\}, \ \ \ C(s,0)=1,
\end{aligned}
\end{equation}
where one has defined an ``ad hoc'' function $A(m,n)$:
\begin{equation}
  \begin{aligned}
&A(m,n) = {{m}\choose{2}} {{m-2}\choose{2}} \cdots {{m-2(n-1)}\choose{2}},
\\
&A(m,n) = 0 \ \ \ [n < 0], \ \ \ A(m,n) = 1 \ \ \ [n=0], \ \ \ A(m,n) =0 \ \ \ [m < 2n].
\end{aligned}
\end{equation}

\end{widetext}

\bibliography{paper}

\end{document}